\documentclass[bibliography=totoc,ngerman,english,11pt,a4paper]{scrartcl}
\pdfoutput=1

\usepackage{jinstpub}
\usepackage[utf8]{inputenc}
\usepackage[T1]{fontenc}
\usepackage[main=english,ngerman]{babel}
\usepackage{textcomp} %
\usepackage{gensymb}  %
\usepackage{graphicx} %
\usepackage{amsmath} %
\usepackage{amssymb}
\usepackage{bm} %
\usepackage{xspace}
\usepackage{url}
\usepackage{hyperref}
\usepackage{sidecap} %
\usepackage{units}
\usepackage{csquotes}
\usepackage[frozencache,cachedir=minteddir]{minted} %
\usepackage{import} %
\usepackage{doi} %
\usepackage[
    backend=bibtex, %
    style=numeric, %
    sorting=none, %
    natbib=true, %
    url=true, %
    doi=true, %
    eprint=true, %
    arxiv=abs %
]{biblatex}
\renewbibmacro{in:}{} %
\newcommand{\Aref}[1]{{%
    \def\chapterautorefname{Chapter}%
    \def\sectionautorefname{Section}%
    \def\subsectionautorefname{Section}%
    \def\subsubsectionautorefname{Section}%
    \def\figureautorefname{Figure}%
    \def\tableautorefname{Table}%
    \def\equationautorefname{Eq.}%
    \autoref{#1}%
}}
\newcommand{\aref}[1]{{%
    \def\chapterautorefname{chapter}%
    \def\sectionautorefname{section}%
    \def\subsectionautorefname{section}%
    \def\subsubsectionautorefname{section}%
    \def\figureautorefname{figure}%
    \def\tableautorefname{table}%
    \def\equationautorefname{eq.}%
    \autoref{#1}%
}}

\newcommand{\tablelead}{\hline\\[-1em]}
\newcommand{\tabletail}{\hline}

\newcommand{\ReMU}{\texttt{ReMU}\xspace}

\newcommand{\lik}{L} %
\newcommand{\dd}{\mathrm{d}} %
\newcommand{\orderof}{\ensuremath{\mathcal{O}}}        %

\title{A response-matrix-centred approach to presenting cross-section measurements}
\author{L.~Koch}
\affiliation{RWTH Aachen University,\\III. Physikalisches Insitut B,\\Aachen, Germany}
\affiliation{STFC Rutherford Appleton Laboratory,\\Particle Physics Department,\\Didcot, United Kingdom}
\emailAdd{lukas.koch@stfc.ac.uk}

\abstract{The current canonical approach to publishing cross-section data is to unfold the reconstructed distributions.
Detector effects like efficiency and smearing are undone mathematically, yielding distributions in true event properties.
This is an ill-posed problem, as even small statistical variations in the reconstructed data can lead to large changes in the unfolded spectra.

This work presents an alternative or complementary approach: the response-matrix-centred forward-folding approach.
It offers a convenient way to forward-fold model expectations in truth space to reconstructed quantities.
These can then be compared to the data directly,
similar to what is usually done with full detector simulations within the experimental collaborations.
For this, the detector response (efficiency and smearing) is parametrised as a matrix.
The effects of the detector on the measurement of a given model is simulated by simply multiplying the binned truth expectation values by this response matrix.

Systematic uncertainties in the detector response are handled by providing a set of matrices according to the prior distribution of the detector properties and marginalising over them.
Background events can be included in the likelihood calculation by giving background events their own bins in truth space.

To facilitate a straight-forward use of response matrices, a new software framework has been developed: the Response Matrix Utilities (\ReMU).
\ReMU is a Python package distributed via the Python Package Index.
It only uses widely available, standard scientific Python libraries and does not depend on any custom experiment-specific software.
It offers all methods needed to build response matrices from Monte Carlo data sets,
use the response matrix to forward-fold truth-level model predictions,
and compare the predictions to real data using Bayesian or frequentist statistical inference.
}


\begin{document}

\maketitle
\flushbottom

\section{Motivation}

\begin{figure}
    \centering
    \includegraphics[width=0.6\textwidth]{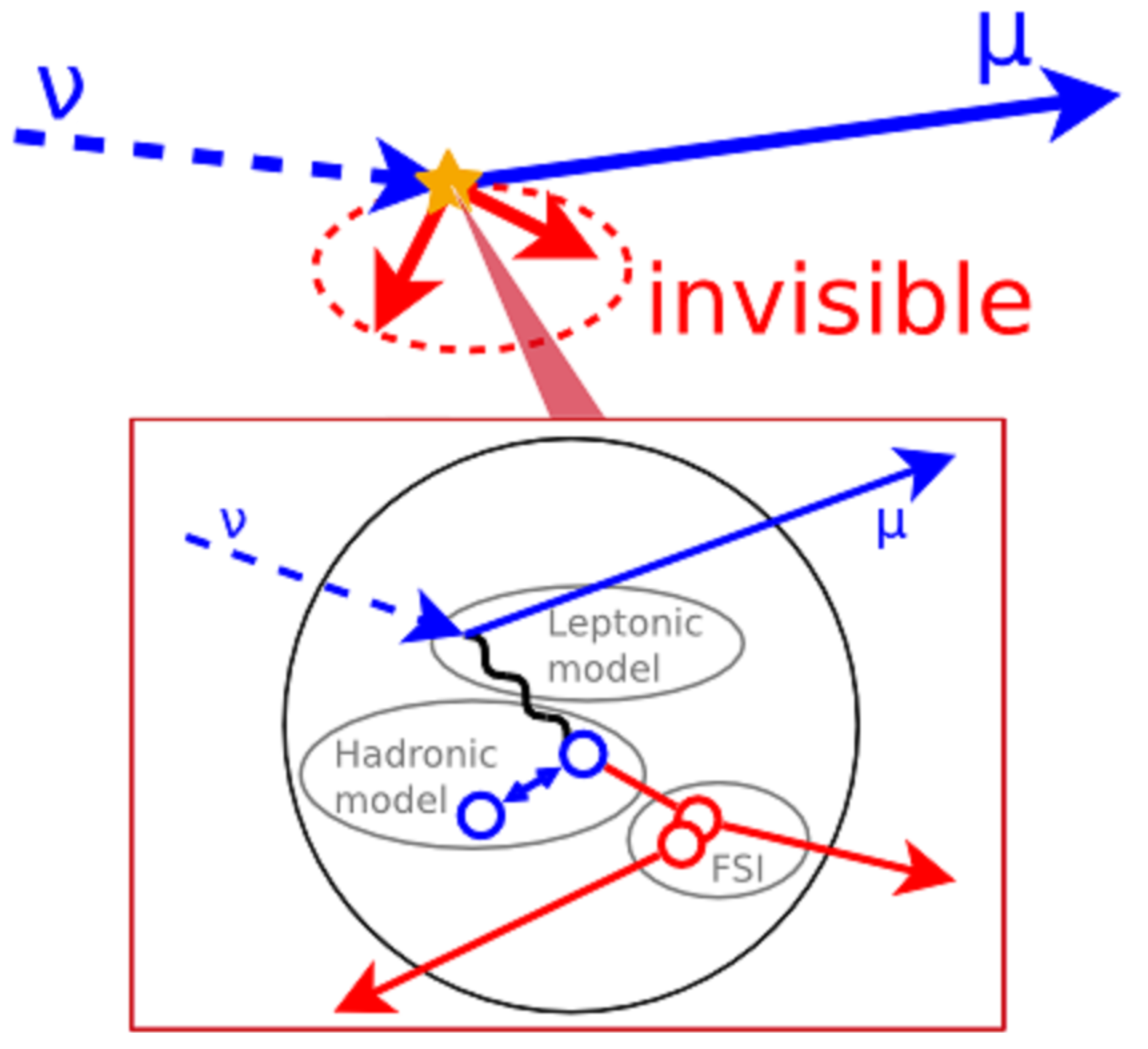}
    \caption[Neutrino cross sections for oscillation experiments]{\label{fig:FSI}%
        Neutrino cross sections for oscillation experiments.
        Neutrino oscillation experiments need to reconstruct the neutrino energy event by event.
        Since the neutrino is invisible, only the products of the interaction can be used for this.
        Exact reconstructions are made impossible by undetectable particles below the detection threshold of the respective detectors.
        Models for electroweak nuclear interactions are used to correct these effects.
        Currently there are quite large theoretical uncertainties on especially the hadronic model (initial state of the nucleus, nucleon form factors, etc.) and the \emph{Final State Interactions} (FSI).
        Cross-section measurements play an important role in constraining these uncertainties (see e.g.~\cite{Alvarez-Ruso2018}).
    }
\end{figure}

Cross-section measurements are an important tool for investigating possible manifestations of \enquote{new physics},
i.e. phenomena beyond the currently accepted models.
This is either done directly with the cross-section measurements,
e.g. in beyond-the-standard-model searches at collider experiments,
or by using cross-section measurements as inputs for other experiments,
e.g. the use of neutrino cross-section measurements for constraining systematic uncertainties in oscillation experiments (see \aref{fig:FSI}).

It is thus important to present these results in a way that allows re-interpretation of the data when new insights into the theoretical models are gained in the future.
Especially in the case of neutrino cross-section measurements it can be difficult to disentangle the measured quantities from detector effects and (possibly poorly motivated) model assumptions (see \cite{Uchida2018} for an overview of challenges and possible solutions).
Measurements that do not take care of these issues can end up being difficult to interpret and ultimately become useless for global data comparisons, fits, etc.

There is no single recipe that ensures that a measurement is free of model assumptions or detector effects.
A couple of points are important to keep in mind though.
For example, when trying to constrain a certain aspect of an electroweak nuclear interaction model,
it is important to make sure that no assumptions of that model are influencing the measurement.
In the worst case, one can end up publishing \enquote{data} that is really just a carbon copy of the model.
Checks against these kinds of model bias are a common part of physics analyses.

It is equally important though, to ensure that (possibly implicit) assumptions about model effects one is \emph{not} interested in do not affect the measurement.
The detection efficiency and reconstruction resolution of an event in a real detector can depend on a lot of variables.
In principle, it depends on the type, momentum and direction of every single particle that leaves a vertex and can theoretically be detected.
In practice, one is interested in the properties of only a few of those particles.
Even when considering just one particle, it is often not fully characterised by its three momentum components,
but the information is reduced to, for example, the magnitude of the momentum to avoid bins with very few events in n-dimensional histograms.

Unfortunately, not looking at the other variables does not mean that their influence on the detector efficiency goes away.
Models that are well tuned to real data in the distribution of a certain quantity, like the total lepton momentum in a charged-current neutrino interaction,
can differ wildly from reality in distributions that simply have not been looked at before, e.g. the second highest proton momentum in an event.
Ignoring these quantities means that one uses the average efficiency of the events,
obtained for a certain distribution of the ignored quantities.

This can lead to very different efficiencies and purities of event selections
if two theories predict very different distributions.
For example, all detectors have certain energy/momentum thresholds below which they are not sensitive to particles.
If two theories (or a theory and reality for that matter) now predict different fractions of events/particles below that threshold,
the resulting average efficiency of selecting the events will vary accordingly.

A lot of work is done on minimising or at least quantifying these effects.
Strategies range from doing multi-dimensional differential cross-section measurements (to ensure all dependencies of the efficiency are modelled),
to repeating the analysis with multiple theories and simply quoting how much the results depend on the used model.
The former approach requires a lot of data to have a significant number of events in every bin,
while the latter suffers from the uncertainty of whether all available models even cover reality at all.

The response-matrix-centred method described in this work aims to combine the model independence of the multi-dimensional approach with the ability to work with low number of events of the naive model test.
This is achieved by de-coupling the binning of the reconstructed events from the description of the events at the generator level.
The high dimensionality of variables is only needed in \emph{truth space}, i.e. the description of the events at the generator level.
The actual recorded data can be binned much coarser in \emph{reco space}, i.e. with wider binning and/or fewer reconstructed variables.
The response matrix is the connecting piece between the two, describing how likely an event in a particular truth space bin is going to end up in any of the reco space bins.

If the truth binning is chosen carefully, the response matrix should be
(sufficiently\footnote{A certain dependence on the event distribution within the bins will always remain, but it can be reduced to the point where it does not matter compared to other uncertainties.})
independent of any assumed physics model of the interactions.
That is, different models can predict different truth space distributions,
but the values of the response matrix elements do not depend on the model that is used to build the matrix\footnote{Aside from statistical effects from the number of available simulated events in each truth bin.}.
The real data and response matrix can then be used with arbitrary models to calculate a likelihood and extract cross sections.

This is so far not different from the naive model testing method.
The advantage of the response matrix approach is realised when considering the matrix and the raw data as the main result of the measurement.
They are (ideally) independent of any model assumptions and can be used to test any new model or model improvement that will be developed in the future.
Furthermore, if the raw data and response matrix are published, model developers can use them directly to test new models against old data.
Compared to the classical approach, where the theories are developed by theorists and then tested within the experimental groups in dedicated analyses,
this reduces the time of the development cycle considerably.
In fact, a lot of work has been spent to make old experimental results available for easy model tuning, for example with the NUISANCE\cite{nuisance} or Rivet\cite{Buckley2010} frameworks.
Results obtained with the response-matrix-centred approach would be very easy to include in such global fits.

It might seem like a shortcut for lazy experimental physicists to simply publish the raw data and response matrix to leave the rest to the model builders.
This is not the case though, since the construction of the response matrix requires exactly the same understanding of the detector and care to cover all systematics as a classical, unfolding analysis.
Also it is unlikely that any experimental group would publish the data and response matrix without also using them for their own model tests.

It is worth noting that model comparisons in reconstructed (or smeared) space are in general more powerful than equivalent model tests in truth (or unfolded) space \cite{Cousins2016}.
The forward-folding approach might thus also be advantageous for analyses that \emph{do} have enough statistics to do a multi-dimensional unfolding of the results.
In any case, one is not restricted to do one or the other.
If the data, time and person-power allow it, it might be the best choice to publish both an unfolded result, as well as the raw data with a response matrix.
The additional work needed for doing an unfolding analysis on top of a forward folding one is probably less than it might seem.
The response matrix can be used to do an unfolding analysis with it,
e.g. using a likelihood fit or Markov Chain Monte Carlo with the bin-by-bin truth-level predictions as fit parameters.
If the model-independence criterion of the forward-folding matrix leads to very underconstrained truth bins
(i.e. a much finer truth binning than in reconstructed space),
the dimensionality can be reduced by fitting templates of a theory.
This would mean the result is no longer model-independent,
but it could be argued that a purely unfolding analysis should suffer from the same problems.

The description of the mathematical model of the response-matrix-centred approach can be found in \aref{sec:strategy}.
Details on how to build the matrix and how to contain the knowledge about the systematic uncertainties in it are given in \aref{sec:implementation}.
The algorithms are implemented in a Python software library called \ReMU, Response Matrix Utilities.
It is intended to make the usage of the data and response matrix as easy as possible.
More informations about the software and data formats are included in \aref{sec:software}.

\section{Measurement strategy}
\label{sec:strategy}

\subsection{Introduction}

The response-matrix-centred approach is a way of presenting cross-section measurements (or any other kind of counting experiment) in a way that tries to be as model-independent as possible.
Its main philosophy can be summarised in three main points:
\begin{enumerate}
    \item There is a linear relationship between \enquote{true} physics expectation values,
        i.e. as described on the generator level,
        and expected number of measured events.
    \item Our knowledge of that relationship is imperfect.
    \item The data is the data is the data.
\end{enumerate}
The linear relationship mentioned in the first point is the response matrix.
It describes how likely it is to count an event that happened in the detector (efficiency)
and in which reconstructed bin it is probably going to end up, i.e. what the reconstructed properties of the event will be (smearing).
We know the elements of this matrix only to a certain precision.
They are subject to uncertainties of evaluating them using Monte Carlo (MC) simulations of events in the detector..

The actually measured data on the other hand is the only thing we can be 100\% sure about.
It consists of exact numbers, and systematic or even statistical errors only apply if one interprets the actual data as expectation values for future measurements.
For example. if we do a cross-section measurement and measure 16 events of a certain type, we measure \emph{exactly} 16 events, not something between 12 and 20.
Once we try to predict future repetitions of the experiment, we have to interpret this number as measurement of the expectation value, so we get an uncertainty on that: the expectation value is $16 \pm 4$.
In general, there is no one-to-one correspondence between data and the physics variables we are interested in,
so the response matrix must be used to translate between the two.

\begin{figure}
    \centering
    \includegraphics[width=0.9\textwidth]{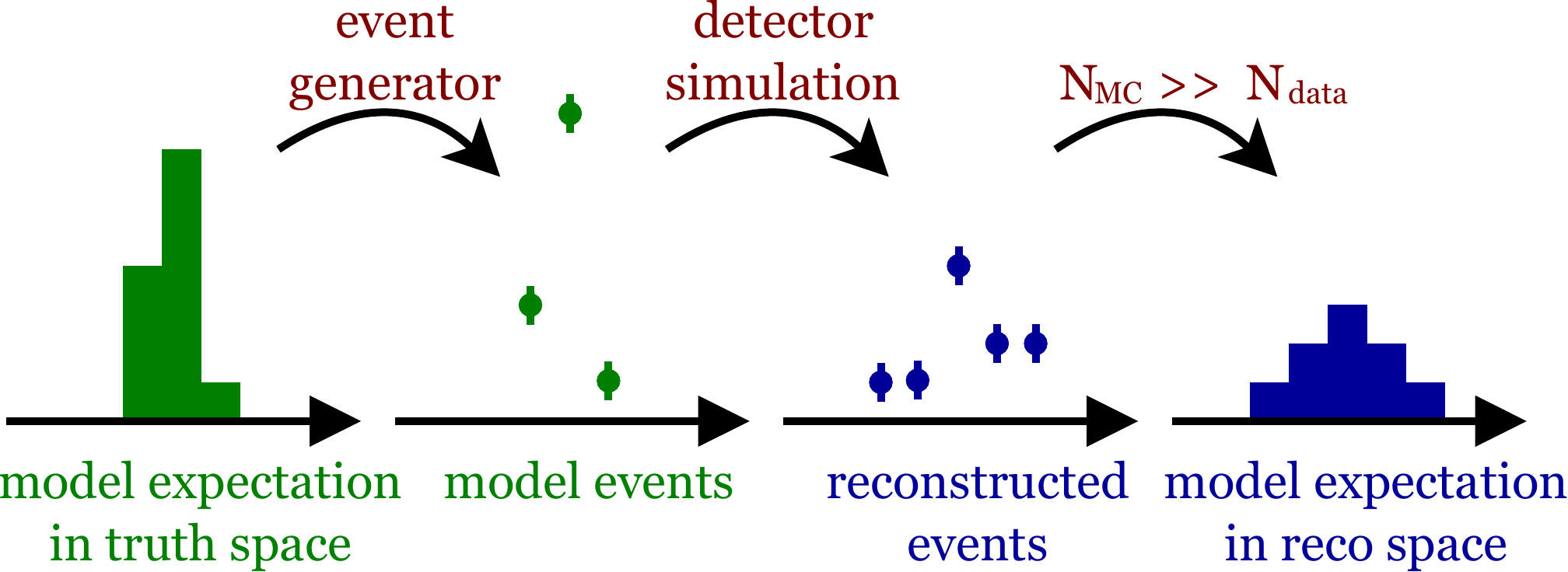}\\[2em]
    \includegraphics[width=0.7\textwidth]{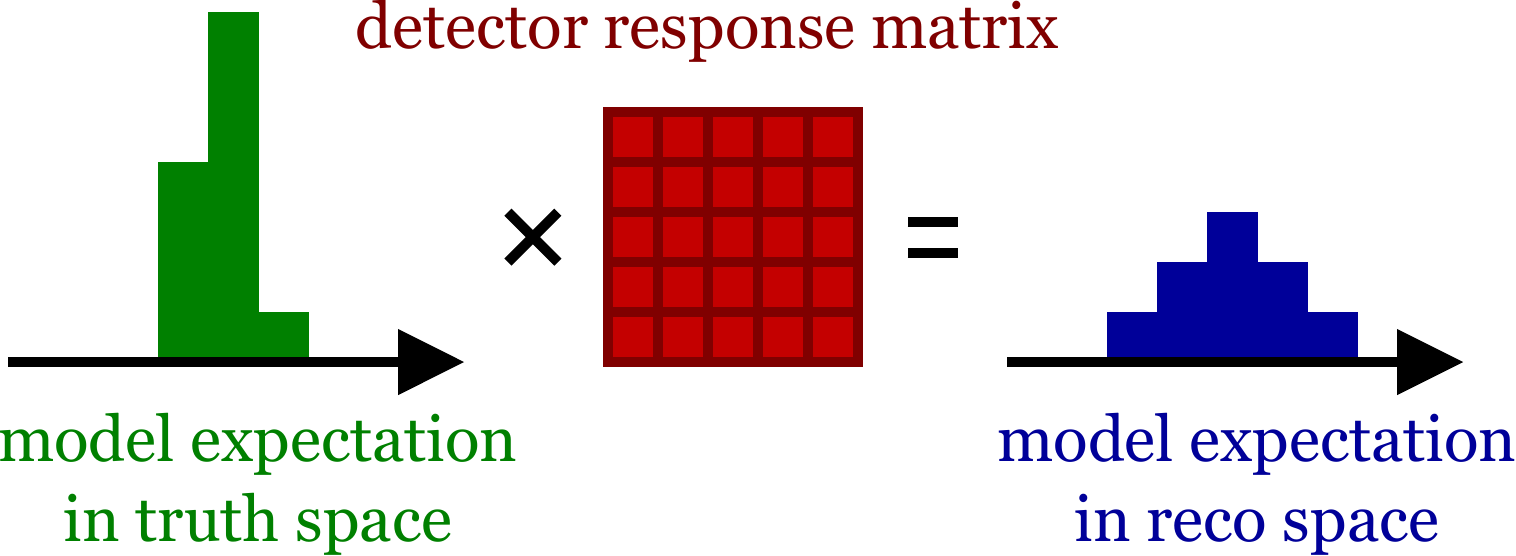}\\[2em]
    \includegraphics[width=0.6\textwidth]{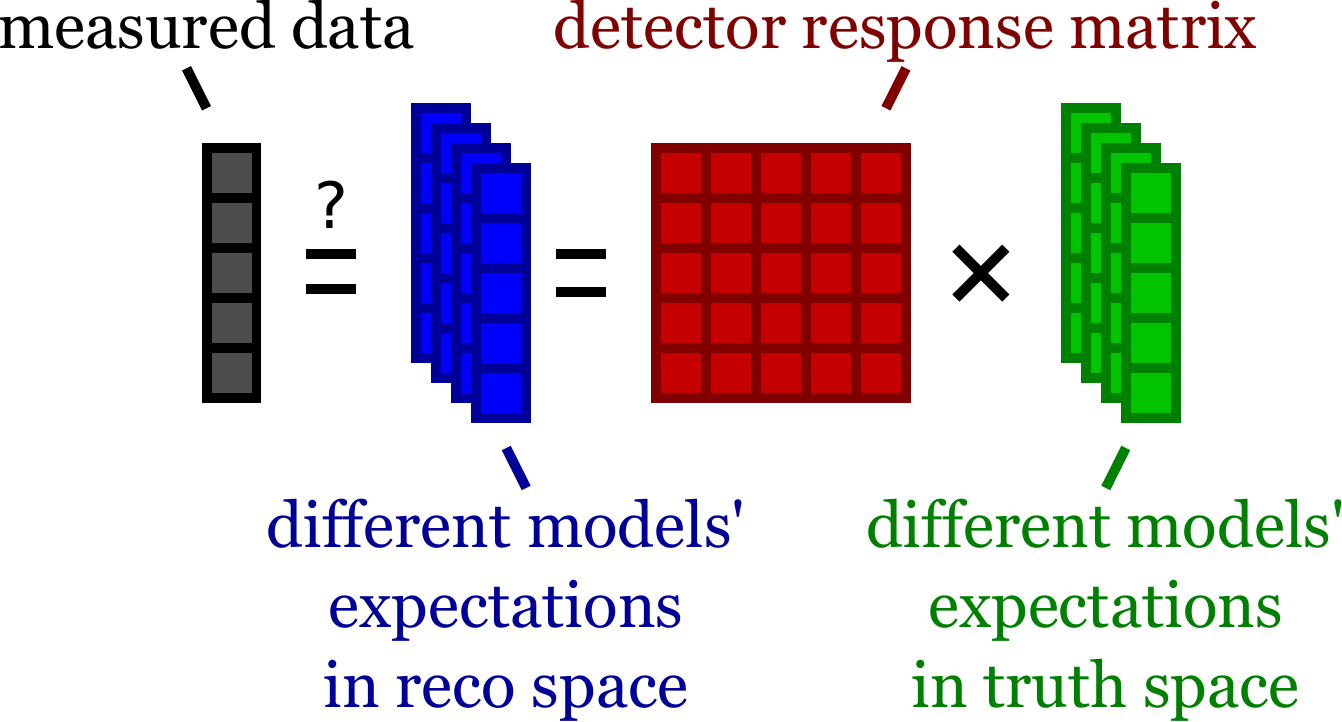}
    \caption[The response-matrix-centred approach]{\label{fig:approach}%
        The response-matrix-centred approach.
        The aim of the approach presented here is to replace the computationally intensive full detector simulation (top)
        with a much simpler matrix multiplication (centre).
        This would allow a much faster test of different models against the data (bottom).
    }
\end{figure}

Ideally, a measurement should remain useful not only for the current interaction model,
but also for all possible future models.
This can be achieved if:
\begin{itemize}
    \item Arbitrary models can be checked for compatibility with the published data.
    \item The publication contains all tools and information to do this.
    \item These tools do not depend on the currently favoured model.
\end{itemize}
All of this is possible with the response-matrix-centred approach (see \aref{fig:approach}).

The idea of bringing truth-space expectation values into the reconstructed space is not new.
This is regularly done in most experiments.
Unfortunately this usually requires expert knowledge and access to the (simulation) software stack of the experiment,
which is often only available to collaborators.
Simplifying the process by providing a response matrix could enable non-collaborators to bring their models into the reco space of the experiment,
as well as speed up the process of testing models within a collaboration.

The main result of any measurement presented in this way consists of the \emph{raw} reconstructed data (without any systematic errors)
and the response matrix including all uncertainties on the matrix elements.
These two objects are everything that is needed to test arbitrary physics models against the data in a consistent way.
The tests then produce \enquote{traditional} results in truth space.

\subsection{The detector model}
\label{sec:detectormodel}

We categorise all events by their true properties and sort them into a set of truth bins.
For simulated events these properties are directly accessible, while for real data they remain hidden.
Selected events are also binned according to their reconstructed properties.
The Poisson expectation value for the number of events in the $j$-th truth bin is~$\mu_j$.
It is determined by the underlying physics models and the experimental setup,
e.g. target material, mass and integrated neutrino flux.

If an event happens in truth bin~$j$, it has a certain probability~$P(j \rightarrow i)$ to be selected and reconstructed in the $i$-th reconstruction bin.
This probability can be calculated from MC samples with known true and reconstructed properties:
\begin{equation}
    P(j \rightarrow i) = \lim_{N \rightarrow \infty} \frac{N(\text{truth}=j,\text{reco}=i)}{N(\text{truth}=j)} \text{.}
\end{equation}
It should depend \emph{only} on the detector properties and \emph{not} on the interaction model.
This can be achieved by choosing an appropriate binning in truth space (see \aref{sec:binning}).
The expectation value for the $i$-th reconstruction bin $\nu_i$ is then
\begin{equation}
    \nu_i = \sum_j P(j \rightarrow i)\mu_j \text{.}
\end{equation}
This can be expressed as a matrix product (using Einstein notation)
\begin{equation}
    \nu_i = R_{ij}\mu_j \text{,}
\end{equation}
where $R$ is the detector response matrix.
Please note that this matrix models both the selection efficiency and reconstruction smearing.

Truth and reconstructed space will generally be binned in multiple variables,
which might give the impression that the response matrix needs to be an n-dimensional object.
This is not the case, as both binnings can be linearised, i.e. assigning each bin an identifying integer.%
\footnote{In fact, the binning does not even have to be \enquote{regular} in any way.
The bins can have arbitrary shapes in arbitrary dimensions.
The only thing that is demanded of the bins is that they do not overlap,
so each event is assigned exactly one truth bin and at most one reco bin.}
The response matrix should now be seen as a regular, two-dimensional matrix that translates between the two one-dimensional vectors of bins.

Since we need to know the truth information, $R$ can only be built from MC samples.
Unfortunately the simulated detector does not mirror the real detector perfectly.
The differences are parametrised in a set of systematic uncertainties, e.g. an uncertainty on the momentum resolution or the track reconstruction efficiency.
Their effect on the response matrix can be evaluated by producing lots of \enquote{toy simulations},%
\footnote{Depending on the collaboration, these are also called \enquote{universes}.}
in which the same dataset is processed, but the detector properties are sampled from their uncertainty distribution.

Because it is often impractical to run the full detector simulation hundreds of times,
these toy simulations are commonly created by modifying a single full \enquote{baseline} simulation.
Depending on the type of systematic uncertainty, this can be done by assigning weights to events,
or by varying the reconstructed properties of the events.
This yields a set of $N_\text{toy}$ response matrices~$R^t$, each describing one possible true detector and its reconstruction expectation values:
\begin{equation}
    \nu_i^t = R_{ij}^t\mu_j \text{.}
\end{equation}

\subsection{The likelihood}
\label{sec:likelihood}

One way to measure the compatibility of a given hypothesis and the measured data is the likelihood~$\lik$.
For a discrete counting experiment, it describes the probability of getting exactly the measured result~$\bm{n}$,
given the tested hypothesis~$\bm{\theta}$:
\begin{equation}
    \lik(\bm{\theta}) = P(\bm{n}|\bm{\theta}) \text{.}
\end{equation}
In our framework, the hypothesis is described by the expectation values of the truth bins $\bm{\mu}$:
\begin{equation}
    \lik(\bm{\mu}) = P(\bm{n}|\bm{\mu}) \text{.}
\end{equation}
We can expand this expression to explicitly include the possibility of different detector responses:
\begin{equation}
    P(\bm{n}|\bm{\mu}) = \int_R P(\bm{n}|\bm{\mu},R) f(R) \dd R \text{,}
\end{equation}
where the integral is over all possible detectors~$R$ and the probability density~$f(R)$ of them being true.
This is impractical, but we can replace the infinite, high-dimensional integral with a random sample of toy detectors~$R^t$:
\begin{equation}
    P(\bm{n}|\bm{\mu}) = \frac{1}{N_\text{toy}} \sum_{t} P(\bm{n}|\bm{\mu},R^t) \text{.}
\end{equation}
The sample is drawn from the uncertainty distributions of the detector properties $f$,
so more-probable matrices will appear more often than unlikely ones.
Within the set of toy matrices, each one is equally likely.

The remaining probability term is just that of a multi-bin Poisson counting experiment:
\begin{align}
     P(\bm{n}|\bm{\mu},R^t) &= P_\text{Poisson}(\bm{n}|\bm{\nu}=R^t\cdot\bm{\mu}) \nonumber \\
                            &= \prod_i \frac{(R^t_{ij}\mu_j)^{n_i}}{n_i!}\exp(-R^t_{ij}\mu_j)
\end{align}
So ultimately the total marginal likelihood of a tested hypothesis, given the measured data, is
\begin{equation}
    \lik(\bm{\mu}) = P(\bm{n}|\bm{\mu}) = \frac{1}{N_\text{toy}} \sum_{t} \prod_i \frac{(R^t_{ij}\mu_j)^{n_i}}{n_i!}\exp(-R^t_{ij}\mu_j) \text{.}
\end{equation}
Alternatively one can also choose to use the profile likelihood
\begin{equation}
    \lik_\text{profile}(\bm{\mu}) = \max_{t} \prod_i \frac{(R^t_{ij}\mu_j)^{n_i}}{n_i!}\exp(-R^t_{ij}\mu_j) \text{,}
\end{equation}
which just selects the toy migration matrix with the highest resulting likelihood.

\begin{figure}
    \centering
    \includegraphics[width=0.8\textwidth]{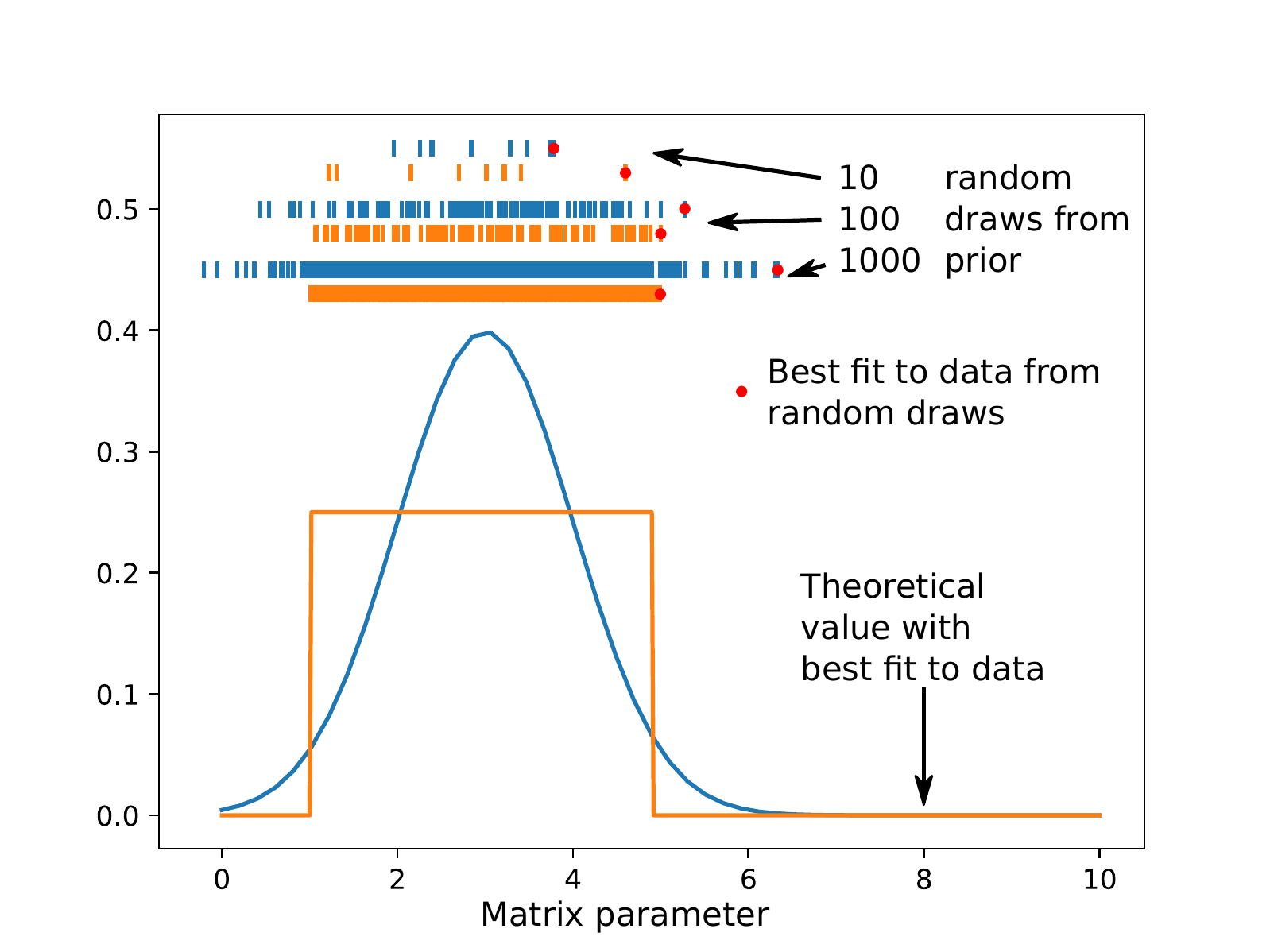}
    \caption[Profile likelihood and unbounded detector parameters]{\label{fig:profile-problem}%
        Profile likelihood and unbounded detector parameters.
        If the best possible fit value for a detector parameter lies outside the expected range,
        this can led to unwanted effects in combination with the use of a profile likelihood.
        The best fit value (red) keeps increasing with the number of random evaluations in the case of the normally distributed parameter (blue).
        When the parameter uncertainty is assumed to be a bound uniform distribution, the value approaches a limiting value much quicker (orange).
    }
\end{figure}

Using the profile likelihood with a discrete set of toy matrices can lead to results
that are very dependent on the number of simulated toys,
e.g. when the parameters of the systematics are not strictly bound.
If a parameter of the matrix is distributed without strict limits%
\footnote{E.g. with a normal distribution, which is not bounded in either direction.}
and the maximum likelihood is achieved for very extreme matrices,
the achieved likelihood will depend a lot on the number of toy matrices (see \aref{fig:profile-problem}).
The more matrices are sampled from the unlimited distribution, the more extreme the most extreme matrix will become.
If, on the other hand, all parameters are sampled from bounded distributions\footnote{E.g. uniform distributions.},
the extremeness of the most extreme matrix will tend to a limiting value instead of rising towards infinity with the number of toy matrices.

Likelihoods calculated with the response matrix can then be used in standard frequentist or Bayesian inference methods.
Some examples and explanations are given in \aref{sec:hypothesis-tests} and onwards.

\subsection{Backgrounds}

In general, data will contain background events that are not part of the process one wants to investigate.
Within the forward-folding approach, there are three ways to deal with these background events.
Simply subtracting them from the data vector is not an option, even if the background contamination is perfectly known.
This would break the Poissonian assumptions in the likelihood function.

We can roughly divide the background in three categories:
\begin{enumerate}
    \item irreducible background,
    \item \enquote{physics-like} background,
    \item detector specific background.
\end{enumerate}
Irreducible background produces exactly the same (measurable) signal in the detector as signal events.
Since they are identical, as far as the detector is concerned, they occupy the same truth bins,
and it is not possible to tell them apart on an event-by-event basis.
The signal purity within a truth bin is always determined by the tested models,
since those also determine any connections between different truth bins (background shapes, etc.).
Please note the distinction between the definition of the truth bins
-- which is dictated by what can be measured with the detector --
and the definition of the \enquote{signal of interest},
which depends on the tested models.

\begin{figure}
    \centering
    \includegraphics[width=0.8\textwidth]{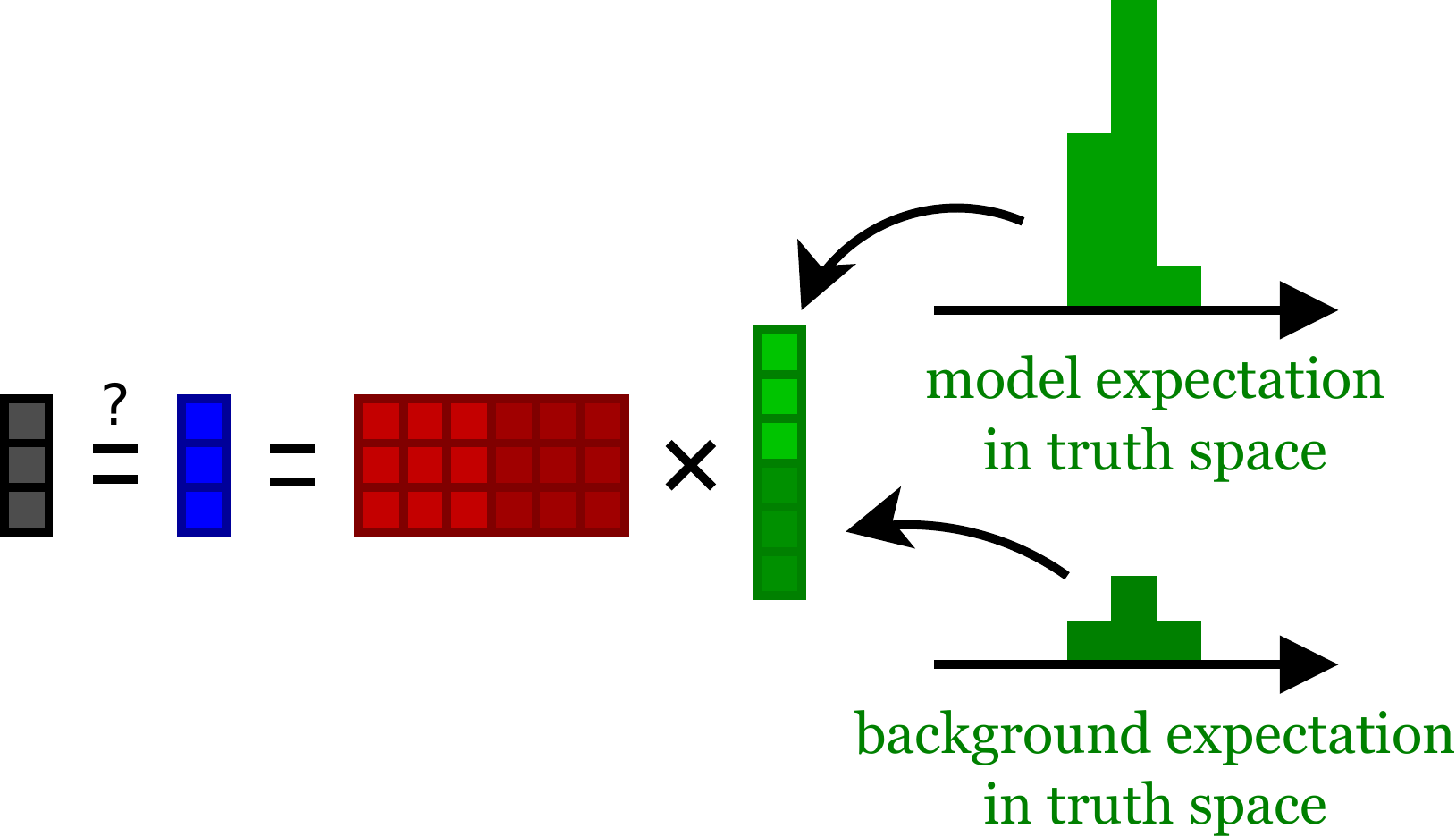}
    \caption[Forward-folded background]{\label{fig:folded-BG}%
        Forward-folded background.
        Background processes get their own separated binning in truth space.
        Future changes in the modeling of the background are possible.
        Data releases can include templates of the background distributions,
        so users of the response matrix will not have to provide their own background estimates.
    }
\end{figure}

\enquote{Physics-like} background does in principle produce detector signatures that are different from the signal.
When it ends up in the final selection, it is usually due to some sort of reconstruction failure,
like a misidentified or missed particle.
The shape and amount of physics-like background depends on physics models that might be not well constrained at the moment,
so it should be treated like the signal, with its own bins in truth space and response matrix columns (see \aref{fig:folded-BG}).
These truth bins are \emph{separate} from the truth bins of the signal events.
Since the physics-like backgrounds produce different kinds of events,
the detector response to these events is also different from the response to the signal events.
It can be seen as its own response matrix, but the toy variations have to be consistent with those of the signal events.
This allows the data to be used consistently with evolving signal and background physics model predictions.
Since users of the data and response matrix release might not care about the background model,
the publication can include such models, e.g. as templates in truth space.

\begin{figure}
    \centering
    \includegraphics[width=0.8\textwidth]{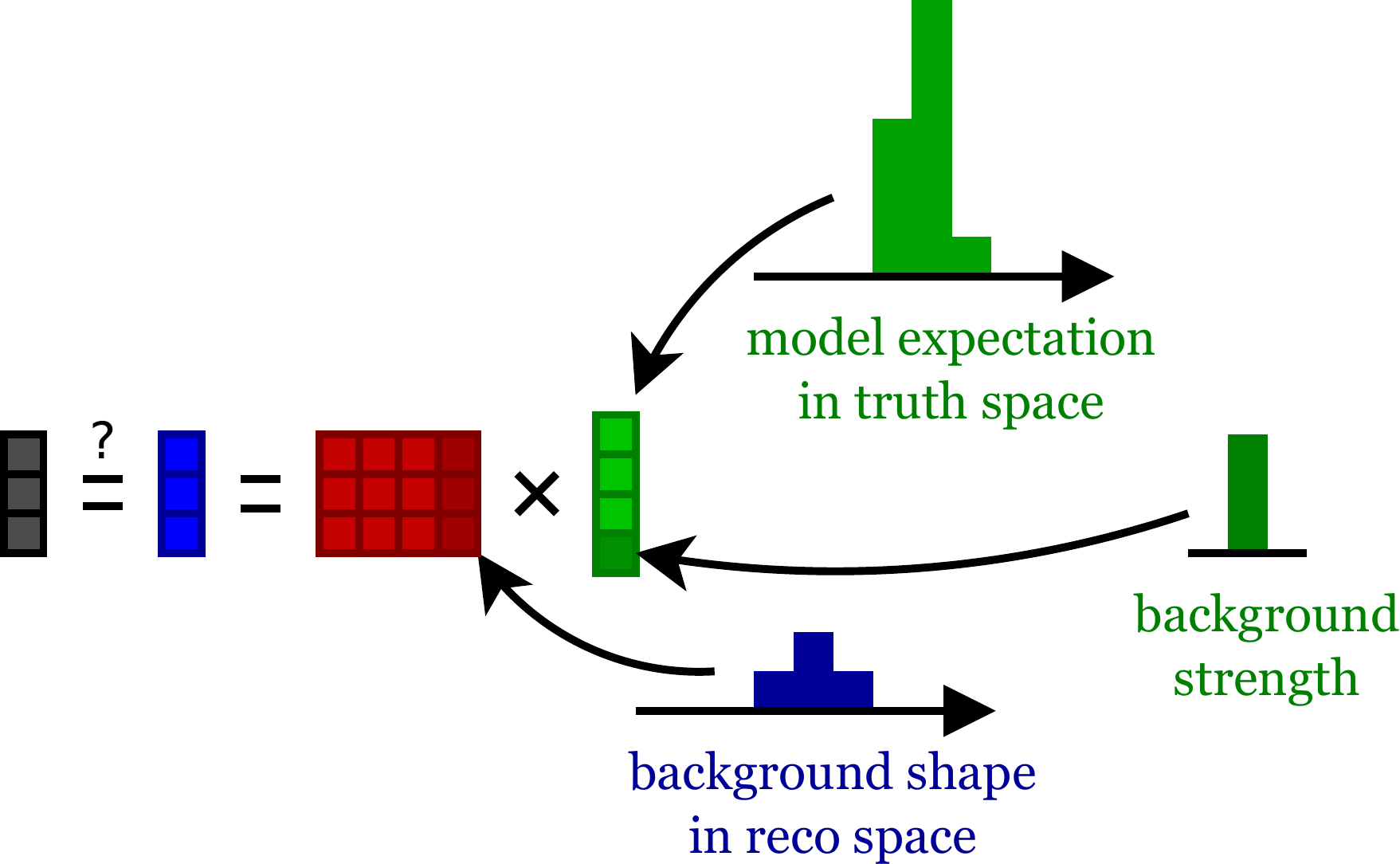}
    \caption[Template background]{\label{fig:template-BG}%
        Template background.
        The reco-space shape of the background is stored as columns in the response matrix.
        The corresponding truth bins decide the strength/weight of the background.
        Future changes in the modeling of the background are \emph{not} possible.
        Only the weights can be varied.
    }
\end{figure}

If the background is very detector specific and does not depend (much) on unconstrained physics models,
it could be simpler to encode the reconstructed background shape in the columns of the response matrix (see \aref{fig:template-BG}).
Each such column corresponds to a single bin in truth space,
which would then decide the strength/amount of that background in the sample.
This would make it impossible to change the background shape in the future,
but it can reduce the complexity of the response matrix considerably.

For both physics-like and detector specific background it is possible to give the data the power to constrain their contributions.
This can be done by including one or more control regions in the reco data vector,
each being enriched in different types of background events.
Reco-space model predictions in the control regions and the signal region are correlated by the response matrix,
but the data points remain statistically independent Poissonian samples.

\subsection{Matrix tests}
\label{sec:matrix_tests}

All methods described in this work depend on the model-independence of the response matrix,
so this needs to be ensured with dedicated tests.
In general, no response matrix will be completely model-independent,
but the aim is to reduce the model dependence to a level where it can be neglected.
The easiest way to test the model dependence of the response matrix
is to use multiple different event generators, i.e. models, to generate the matrix
and then compare the matrices with one another.

All matrices are only known to a certain precision.
This uncertainty is expressed as a Bayesian posterior probability distribution of the true matrix parameters (see \aref{sec:implementation}).
For the purpose of the comparison of two matrices, we can interpret the set of matrix elements~$R_{ij}$ as a multivariate random variable~$\bm{R}$.
The two posterior distributions of the compared matrices ($\bm{R}$ and~$\bm{R}'$) define a combined distribution of matrix differences~$\bm{X}$:
\begin{equation}
    \bm{X} = \bm{R} - \bm{R}'
\end{equation}
The two matrices can be considered compatible with one another if the point signifying that the matrices are identical $\bm{X}^0 = (0,0,\ldots)$, i.e. $R_{ij} = R'_{ij}\ \forall i,j$, is a reasonable part of the distribution of~$\bm{X}$.

A way to test this, is using the Mahalanobis distance of~$\bm{X}^0$.
The Mahalanobis distance~$D_M$ is a generalised standard score%
\footnote{I.e. the distance of a point from the centre of a distribution, measured in standard deviations.}
for correlated, multivariate random variables~\cite{Mahalanobis1936}:
\begin{equation}
    D_M(\bm{X}) = \sqrt{ (\bm{X} - E[\bm{X}])^T \cdot \Sigma^{-1} \cdot (\bm{X} - E[\bm{X}]) }\text{,}
\end{equation}
where $\Sigma^{-1}$ is the inverted covariance matrix of the distribution of~$\bm{X}$.
It can be calculated from a randomly generated set of matrices according to their uncertainties (see \aref{sec:implementation}).
The higher the Mahalanobis distance, the more unlikely a point is under the assumption of the given random distribution.%
\footnote{Assuming a sufficiently symmetric and unimodal distribution.}

The \enquote{extremeness} of~$\bm{X}^0$ can be measured by the probability included in the area of the distribution where $D_M(\bm{X}) < D_M(\bm{X}^0)$.
Equivalently one can define the \enquote{compatibility}~$C$ of the two matrices as the probability included in the area where $D_M(\bm{X}) > D_M(\bm{X}^0)$.
In the case of normally distributed~$\bm{X}$, the squared Mahalanobis distance is distributed like a chi-squared distribution: $D^2_M(\bm{X}) \sim \chi^2_k$,
with the number of degrees of freedom~$k$ being the number of response matrix elements.
This means the compatibility can be calculated as:
\begin{equation}
    C = 1 - F(D^2_M(\bm{X}^0), k)\text{,}
\end{equation}
with the cumulative chi-squared distribution function~$F$.
Since it is not given that the differences of the two matrices are normally distributed,%
\footnote{Especially for bins with only very few entries.}
the more general approach is to numerically integrate the probability:
\begin{equation}
    C = \int_{D_M(\bm{X}) > D_M(\bm{X}^0)} f(\bm{X}) \dd \bm{X}\text{,}
\end{equation}
with the probability density function~$f$.

A high compatibility~$C$ means that it is reasonable to assume that the two matrices describe the same true matrix \emph{within their statistical uncertainties}.
Deciding on a critical value for~$C$, below which the two matrices are considered to be too different, is not straight forward.
At 0.5, the matrices are ensured to be identical more likely than not.
This is rather conservative though, as it corresponds to a difference between the two matrices of \enquote{less than $1\sigma$} in the one-dimensional case.
If one is to decide whether to combine two independent measurements of the same variable into one weighted average,
it seems like one would (in general) still combine measurements that differ up to $2\sigma$ (or even $3\sigma$).
This corresponds to a critical~$C$ of 0.05 (0.003).
Since the model-independence is so important for this analysis method though,
and a modification of the truth binning is a relatively cheap operation,
it might be justified to be conservative here.

It is important to note that passing this check is a necessary condition for model-independence, but not a sufficient one.
The available models might not be different enough to reveal any hidden model dependencies given the available MC statistics.
Ideally, one should also test the generation of the response matrix with completely random input vectors for the detector simulation.
That is, instead of using a physics model to generate the events, one uses particle guns with varying underlying event distributions.
If such data is not available with sufficient statistics,
the next best thing to step outside the bounds of the available models is to re-weight those models with arbitrary weight functions.
Of course, in that case the usual limitation of re-weighting applies,
i.e. that only parts of the phase-space that have been simulated to begin with can be re-weighted.

No matter what is done, it remains impossible to prove conclusively that the response matrix is truly model-independent.
In the end, the analyst will have to decide at which point enough has been done to show that the possible remaining model dependence is small enough for the given data.

\section{Implementation}
\label{sec:implementation}

\subsection{Building the detector response matrices}
\label{sec:build}

The detector response matrix is built from Monte Carlo simulations.
Events are first categorised by their truth information and assigned a truth bin number $j$.
Then events that end up being selected and thus have reconstructed properties get assigned a reco bin number $i$.

The probability for an event in truth bin $j$ ending up in reco bin $i$ is
\begin{equation}
    P(j \rightarrow i) = \lim_{N_j \rightarrow \infty} \frac{N_{ij}}{N_j} \text{,}
\end{equation}
as defined in \aref{sec:strategy}.
Here $N_j$ is the number of events in truth bin $j$, including the events that do not get assigned a reco bin,
and $N_{ij}$ the number of events in truth bin $j$ and reco bin $i$.
Since the number of Monte Carlo events is limited by the available computing resources, this value can only be approximated:
\begin{equation}
    R_{ij} = \frac{N_{ij}}{N_j} \text{.}
\end{equation}

The simulated detector does not reproduce the behaviour of the real one perfectly.
We parametrise the estimated difference as a set of systematic uncertainties that get propagated as weights and variations in the selection (see \aref{sec:detectormodel}).
This event weighting and variation is done as a replacement for full simulations of varied detectors to save computing time.
The systematic parameters are sampled from their assumed distributions (e.g. normal or uniform) and the events are weighted and varied accordingly.
We call each sampling of the parameter space a toy simulation.
Each toy simulation $t$ yields its own response matrix
\begin{equation}
    R^t_{ij} = \frac{W^t_{ij}}{W_j} = \frac{N^t_{ij} w^t_{ij}}{N_j w_j} \text{,}
\end{equation}
where $W_j$ and $W^t_{ij}$ are the sum of weights,
and $w_j$ and $w^t_{ij}$ the average weights of all events in the respective bins.
Since the detector variations do not affect the events on the generator level,
the sum of weights in the truth bin $W_j$ is not affected by the toys.

Another important uncertainty in the matrix comes from the statistical uncertainty due to the finite number of simulated MC events.
This means the values of $R^t_{ij}$ will also suffer from statistical variations from the true MC value.
These fluctuations are not represented in the systematic toys, as those do not vary the generated events.
To effectively incorporate this effect in a coherent way,
we decompose the effects of efficiency and smearing in the matrix and build a model that we can draw toy matrices from.
We estimate the statistical uncertainties in a \enquote{Bayesian inspired} three-step process.

The first two uncertainties stem from the multinomial sampling of $N^t_{ij}$.
For the purpose of statistical error estimation, we split the multinomial process in two parts:
\begin{itemize}
    \item A binomial chance of being reconstructed at all (i.e. efficiency) $\epsilon^t_j$
    \item A multinomial probability of ending up in a certain reco bin (i.e. smearing) $p^t_{ij}$
\end{itemize}
\begin{align}
    \epsilon^t_j &= \lim_{N_j \rightarrow \infty} \frac{\sum_i N^t_{ij}}{N_j} \text{,} \\
    p^t_{ij} &= \lim_{N_j \rightarrow \infty} \frac{N^t_{ij}}{\epsilon^t_j N_j} \text{,} \\
    \epsilon^t_j \cdot p^t_{ij} &= \lim_{N_j \rightarrow \infty} \frac{N^t_{ij}}{N_j} \text{.}
\end{align}
We do not know the true values of these parameters, so we can treat them as Bayesian random variables.
Treating the efficiency separately from the smearing makes it easier to find fitting prior parameters for the distributions (see below).

If we assume a beta distribution%
\footnote{The beta distribution is the conjugate prior for binomial distributed likelihoods. See ~\cite{Fink-1997}.}
as a prior for the distribution of $\epsilon^t_j$,
we can use the simulated number of events directly to update the parameters of the prior, $\beta'_{*j}$ and $\beta'_{\dagger j}$, to get the parameters of the posterior:%
\footnote{Usually the parameters of the beta function are denoted as $\alpha$ and $\beta$.
To avoid confusion with the parameters of the Dirichlet distribution $\alpha_{i}$, we decided to use $\beta_*$ and $\beta_\dagger$ respectively instead.}
\begin{align}
    \epsilon^t_j &\sim \mathrm{Beta}(\beta^t_{*j}, \beta^t_{\dagger j}) \text{,} \\
    \beta^t_{*j} &= \beta'_{*j} + \sum_i N^t_{ij} \nonumber \\
        &= \beta'_{*j} + N^t_{*j} \text{,} \\
    \beta^t_{\dagger j} &= \beta'_{\dagger j} + (N_j - \sum_i N^t_{ij}) \nonumber \\
        &= \beta'_{\dagger j} + N^t_{\dagger j} \text{.}
\end{align}
Here $N^t_{*j}$ is the number of selected and $N^t_{\dagger j}$ the number of \enquote{lost}, i.e. not selected, events in truth bin $j$.

We can do the same for the smearing uncertainty if we assume a Dirichlet distribution%
\footnote{The Dirichlet distribution is the conjugate prior for multinomial distributed likelihoods. See ~\cite{Fink-1997}.}
as a prior for the distribution of $p^t_{ij}$.
Again we can use the simulated number of events directly to update the prior's parameters $\alpha'_{ij}$:
\begin{align}
    \bm{p}^t_j &\sim \mathrm{Dir}(\bm\alpha^t_{j}) \text{,} \\
    \alpha^t_{ij} &= \alpha'_{ij} + N^t_{ij} \text{.}
\end{align}

The variances of the resulting posterior distributions are
\begin{align}
    \sigma^2(\epsilon_j) &= \frac{\beta^t_{*j}\beta^t_{\dagger j}} {(\beta^t_{*j} + \beta^t_{\dagger j})^2(\beta^t_{*j} + \beta^t_{\dagger j} + 1)} \text{,} \\
    \sigma^2(p^t_{ij}) &= \frac{\alpha^t_{ij}(\sum_{i'\neq i} \alpha^t_{i'j})} {(\sum_{i'} \alpha^t_{i'j})^2((\sum_{i'} \alpha^t_{i'j}) + 1)} \text{,}
\end{align}
and the expectation values
\begin{align}
    \hat{\epsilon}^t_j &= \frac{\beta^t_{*j}}{\beta^t_{*j} + \beta^t_{\dagger j}} \nonumber \\
        &= \frac{\beta^t_{*j}}{N_j + \beta'_{*j} + \beta'_{\dagger j}} \text{,} \\
    \hat{p}^t_{ij} &= \frac{\alpha^t_{ij}}{\sum_{i'} \alpha^t_{i'j}} \text{.}
\end{align}
As prior parameters we set
\begin{equation}\label{eq:betaprior}
    \beta'_{*j} = \beta'_{\dagger j} = 1
\end{equation}
and
\begin{equation}\label{eq:alphaprior}
    \alpha'_{ij} = \min(1, 3^{N_\text{reco variables}}/N_\text{reco bins})\text{.}
\end{equation}
The choice of prior parameters in \aref{eq:betaprior} ensures that the overall reconstruction (in)efficiency of the truth bins is uniformly distributed apriori.
The prior parameters in \aref{eq:alphaprior} assume that the reconstruction probabilities are concentrated on a few reco bins ($\sim 3$ per reco variable),
while being completely agnostic about \emph{which} reco bins those are.%
\footnote{Dirichlet distributions with $\alpha < 1$ favour \enquote{extreme} sets of $p$,
where most of the probability is concentrated in few categories,
over flat sets, where the probability is more uniformly distributed.
The corresponding reco bins do \emph{not} have to be contiguous.}
Please note that a flat Dirichlet prior ($\alpha'_{ij} = 1$) is generally not suitable for the description of event smearing with many reco bins.
It biases the matrix towards very strong smearing (all truth bins are smeared to all reco bins equivalently in the prior).

The resulting variances of the posterior distributions are consistent with the standard frequentist approach in the limit of high statistics.
Especially the binomial case corresponds to an experiment where we added a \enquote{pseudo-observation} of two simulated events,
of which one was successfully reconstructed, to the actual data.

The third step is to evaluate the statistical uncertainty of the weight correction.
The true weight correction
\begin{equation}
    m^t_{ij} = \lim_{N_j \rightarrow \infty} \frac{w^t_{ij}}{w_j}
\end{equation}
is estimated from the sum of of weights:
\begin{equation}
    \hat{m}^t_{ij} = \frac{w^t_{ij}}{w_j}
                   = \frac{W^t_{ij}/N^t_{ij}}{W_j/N_j} \text{.}
\end{equation}
For the purpose of the variance estimation, we treat $w^t_{ij}$ \emph{independently} from $\epsilon^t$ and $p^t_{ij}$ as arithmetic means of samples with given sizes.

We apply the usual standard error of the mean formula for each average weight.
The sample variance is estimated from the sum of squared weights.
To be able to estimate variances even for bins with only one entry,
we add a pseudo-observation event with an expected weight of 1:
\begin{equation}
    \sigma^2(w^t_{ij}) = \left(\left(\frac{V^t_{ij}+1^2}{N^t_{ij}+1}\right) - \left(\frac{W^t_{ij}+1}{N^t_{ij}+1}\right)^2\right) \frac{1}{N^t_{ij} + 1} \text{,}
\end{equation}
where $V^t_{ij}$ are the sums of the squared weights in the respective bins.
The pseudo-observation represents our prior knowledge of the weights and has no effect in the limit of high statistics.
The variance of the weight correction is then
\begin{equation}
    \sigma^2(m^t_{ij}) = \frac{\sigma^2(w^t_{ij})}{(w_j)^2} + \left(\frac{w^t_{ij}}{(w_j)^2}\right)^2 \sigma^2(w_j) \text{.}
\end{equation}
Here $\sigma^2(w_j)$ is the statistical uncertainty on the average weight in truth bin $j$,
defined analogously to $\sigma^2(w^t_{ij})$.

All that is left now, is to combine the variances of the multinomial sampling and the weight correction:
\begin{equation}
    \sigma^2_\text{MC stat}(R^t_{ij}) = (\hat{\epsilon}^t_j \hat{m}^t_{ij})^2 \sigma^2(p^t_{ij})
                                       + (\hat{\epsilon}^t_j \hat{p}^t_{ij})^2 \sigma^2(m^t_{ij})
                                       + (\hat{p}^t_{ij} \hat{m}^t_{ij})^2 \sigma^2(\epsilon^t_j)
\end{equation}
If the statistical variance is much smaller than the systematic detector variation,
\begin{equation}
    \sigma^2_\text{MC stat}(R^t_{ij}) \ll \sigma^2_\text{syst}(R_{ij}) \approx \frac{1}{N_\text{toy} - 1} \sum_t (R^t_{ij} - \bar{R}_{ij})^2 \text{,}
\end{equation}
for all toy experiments, we can neglect it.
In practice there will almost certainly be bins where this is not the case, e.g. (almost) empty matrix elements.

To deal with these non-negligible statistical uncertainties,
we generate random toy matrices from every systematic toy matrix according to the three step process described above:
First we draw a set of efficiencies and multinomial probabilities from the posterior beta/Dirichlet distributions,
and then we modify these with weight factors calculated from normal distributed mean weights.
\begin{align}
    \epsilon^{t*}_j &\sim \mathrm{Beta}(\beta^t_{*j}, \beta^t_{\dagger j}) \text{,} \\
    \bm{p}^{t*}_j &\sim \mathrm{Dir}(\bm\alpha^t_{j}) \text{,} \\
    w^{t*}_{ij} &\sim \mathrm{Norm}(w^t_{ij}, \sigma^2(w^t_{ij})) \text{,} \\
    R^{t*}_{ij} &= \frac{w^{t*}_{ij}}{w^{t*}_{j}} \,\epsilon^{t*}_j p^{t*}_{ij} \text{.}
\end{align}
These toy matrices are then handled just like the systematic toy matrices to calculate marginal or profile likelihoods.

To further limit the influence of the statistical uncertainties,
we check whether the truth bin expectation values exceed the number of simulated events:
\begin{equation}
    \mu_j \overset{!}{<} N_j \text{.}
\end{equation}
Hypotheses that predict more events in a given truth bin than were simulated
are outside the testable scope of the response matrix.
If the tested hypotheses (e.g. in a likelihood fit or Bayesian posterior sampling) are close to this limit,
it could lead to model dependence of the results.
Therefore it is necessary to check whether this is the case.

\subsection{Binning}
\label{sec:binning}

\subsubsection{General considerations}

The properties of the detector response matrix depend first and foremost on the chosen binning in truth and reco space.
The binning has to balance the following (contradictory) aims:
\begin{itemize}
    \item Ensure the independence of the interaction model $\rightarrow$ large number of truth bins.
    \item Maximise the separation power, i.e. how well well different models can be told apart on the reco level $\rightarrow$ large number of reco bins.
    \item Minimise the influence of statistical errors $\rightarrow$ large number of events per bin.
\end{itemize}
The following sections describe the general methodology of choosing the binning.

\subsubsection{Choosing the variables to bin in}

The response matrix can only be model-independent if it is binned in the right variables.
Variables close to the actual observables are more suited than those that describe the event in a more fundamental way,
which have to be inferred from the measurement.
For example, the lepton momentum of a charged-current neutrino interaction is a good variable,
as the detector can directly measure it.
The neutrino energy on the other hand is a bad choice,
because the translation of neutrino energy to observables in the detector depends on the physics model (FSI, etc).

But even when binning in direct observables only, one has to take care not to introduce hidden model dependencies.
The distribution of events in variables that we do not bin in, might still have an effect on the average detector response.
If different models predict different angular distributions, which in turn change the detector efficiency,
binning only in the muon momentum will \emph{not} be model independent.
One would have to bin in \emph{all} truth variables that affect the detector performance to be truly model independent.
In practice this is not possible, as the available computing power and thus the number of Monte Carlo events is limited.
In any case, it is not necessary to expend lots of time and energy to push the model-dependence to infinitesimally low values
if the measurement is already limited by the amount of real data statistics.
Compromises have to be made.

Aside from detector performance considerations, one of course also has to bin in the variables of interest.
The reco binning is dictated by the physics goals of the measurement.
Again it is important to choose variables as close to the actual observables as possible.
If a variable of interest is the function of other more basic observables,
a binning in those observables would be less susceptible to hidden model dependencies.%
\footnote{A perfect truth binning would of course prevent any model dependencies.}
Unfortunately the number of events per bin decreases exponentially with the number of binning dimensions.

\subsubsection{Bin widths}

As seen in \aref{sec:build}, the efficiencies of the truth bins are estimated -- and toy matrices generated -- using a Bayesian approach
that adds two pseudo-measurements as prior information.
In order to not be biased too much towards that prior, we would like the actual number of events per truth bin to be much larger than the number of pseudo observation.
If the models used for the matrix building are similar to the models that will be tested with the matrix,
the average number of events per truth bin is a good measure for this.
If the building models and tested models vary widely,
it might be better to use another number as figure of merit.
The median number of events per bin could be used, or even a lower percentile.
The exact details of this are not that important, as the number is merely a guideline to use when optimising the binning.
The properties of the matrix will be tested independently of this anyway.
For now, let us demand that $\mathrm{mean}(N_j) \overset{!}{>} 50$.

To maximise the number of events per bin, one could choose a very wide binning.
This can lead to model dependences though, if the detector performance varies considerably within a truth bin.
Since model independence is a primary goal of this analysis, this defines an upper limit for the truth bin sizes.

We estimate the response variation within one bin $\Delta R_{ij}$ from the variation between neighbouring bins:
\begin{equation}
    \Delta R_{ij} = \max_{j'} | R_{ij} - R_{ij'} | \text{,}
\end{equation}
with the neighbouring bins $j'$.
Ideally, one would like this variation to be not much higher%
\footnote{Ideally one would like the variation within the bins to be lower than the statistical uncertainty,
but if there is no actual in-bin variation, the statistical uncertainty will dominate this estimate.}
than the uncertainties on the matrix elements:
\begin{equation}
    \Delta^{\!\prime} R_{ij} = \max_{j'} \left| \frac{R_{ij} - R_{ij'}}{\sqrt{\sigma^2(R_{ij}) + \sigma^2(R_{ij'})}} \right|
                 \overset{!}{<} l \text{,}
\end{equation}
with the normalised in-bin variation $\Delta^{\!\prime} R_{ij}$ and a limit $l \sim \orderof(5)$.
Unfortunately this aim is contradictory to the need to fill each truth bin with sufficiently many events to reduce the influence of the priors (see above).
Also, small scale variations might be hidden within the bins,
so care has to be taken on a variable by variable base to optimise the binning.

If the detector response is sufficiently flat,
the truth bin widths should be adjusted to include the necessary MC statistics.
Other than that, they should be made as small as possible.
The reco bin width is mostly dictated by the physics goals of the analysis, data and MC statistics, and the resolution of the detector.
If the truth binning is chosen in a way to ensure model independence, no reco binning will introduce additional model dependence.
A fine reco binning might expose model dependencies, but the cause is solely in the truth binning.
Conversely, a coarse binning can hide dependencies, so one should aim for as fine a binning as MC and data statistics permit.
This also ensures the best hypothesis testing power.
Reco bins should not be finer than their truth counterparts.

One possible algorithm to decide on the final bin widths is as follows:
\begin{enumerate}
    \item Set reco binning according to expected statistics and physics goals.
    \item Set truth binning very fine.
    \item Merge truth bins until $\mathrm{mean}(N_j) \overset{!}{>} 50$
    \begin{enumerate}
        \item Set limit for in-bin variation $l$.
        \item Merge neighbouring bins with lowest number of entries until limit is reached.
        \item Merge neighbouring bins with lowest in-bin variation $\Delta^{\!\prime} R_{ij}$ until limit is reached.
        \item If necessary, increase $l$ and repeat.
    \end{enumerate}
    \item Fine-tune binning by hand.
\end{enumerate}
After this, the resulting matrix must be checked for sufficient model-independence (see \aref{sec:matrix_tests}).
If it fails, the binning has to be adjusted.

\subsubsection{Empty bins}
\label{sec:empty-bins}

The Monte Carlo samples used to generate the response matrix use a physics model $\bm\mu'$.
In that model, certain areas of the truth phase space are very unlikely to be realised
and the corresponding truth bins will not be filled with a sufficient number of events during the response matrix construction.
This means that we have not enough information about how the detector would react to these kinds of events.
Ideally one would like to build the response matrix with simulation data that covers all possible phase space,
but this is computationally difficult.

Since we cannot predict how those events behave in the detector,
we remove those bins from the vector of truth expectation values $\bm\mu$.
This is equivalent to setting those expectation values to $0$ in all considered hypotheses,
and reduces the dimensionality of $\bm\mu$.
It means that we \emph{cannot} test hypotheses that predict any events in these bins.

There might also be reconstruction bins that never get filled during the construction of the response matrix.
The expectation value in those bins will be close%
\footnote{It will not be exactly 0 due to the generation of statistically varied matrices as described in \aref{sec:build}.}
to 0 for all possible hypotheses.
Finding events in these bins would necessitate further investigation and possibly the generation of more Monte Carlo data.

To judge how well the simulated data covers the real measurement and tested hypotheses,
we can compare the number of simulated events to the number of measured/predicted events:
\begin{align}
    \xi_{\text{reco},i}(\bm{n}) &= \max_{t} \frac{n_i}{N^t_i} \\
    \xi_{\text{truth},j}(\bm\mu) &= \frac{\mu_j}{N_j} \text{.}
\end{align}
Numbers close to or above one indicate that the simulated phase space is not sufficient and should be extended.
More specifically, $\bm\xi_\text{truth}(\bm\mu)$ indicates how well the given hypothesis $\bm\mu$ is covered by the simulation,
while $\bm\xi_\text{reco}(\bm{n})$ shows whether the actual measurement is covered at all.

\subsection{Software and data format}
\label{sec:software}

A lot of particle physics experiments rely on the ROOT analysis framework developed at CERN~\cite{ROOT} as their main data format for storing event data.
These data containers can be very specific to the experiment and often require specialised software tools to read them.
Those tools are often only available within a collaboration and are not intended for external users.

The ultimate goal of the response-matrix-centred approach is to enable people who are not intimately familiar with the experiment to compare event generators with the measured data.
To this end, it was decided to develop the software that deals with the response matrix independently from collaboration-internal frameworks,
and that can be used for both building response matrices as well as using them to test models against published data.
The result of this effort is the Response Matrix Utilities framework \ReMU{}~\cite{ReMU}.

\ReMU is written in pure Python and thus able to run on any system that supports the scripting language.
Numerical calculations are handled by the NumPy~\cite{NumPy}, SciPy~\cite{SciPy}, and PyMC~\cite{PyMC} packages to take advantage of the performance gains of compiled code.
\ReMU's source code is publicly available on the code-sharing platform GitHub~\cite{GitHub},
and releases of the software are distributed via the Python Package Index (PyPI)~\cite{PyPI}.
This means, installing the framework on systems supporting PyPI can be done with a single command:
\begin{minted}{sh}
    pip install remu
\end{minted}
Data is stored and exchanged with standard file formats.
The binning of the response matrix is saved in YAML files~\cite{YAML}, a text format that is both human readable and easy to parse by machines.
The response matrix is saved as a binary NumPy file.
To save disk and RAM space, the matrix is saved as a \enquote{sparse} matrix, i.e. only the rows of the matrix corresponding to truth bins that were filled during the matrix creation are saved.
The information which bins were filled (and how many events were simulated in each) is saved in another binary NumPy file.
The data itself (reco or truth space) can either be provided as binned histograms with binary NumPy files,
or event-by-event with Comma-Separated Values (CSV) files.
Other file formats are supported via the python data analysis library \enquote{pandas}~\cite{pandas}.
ROOT files can be read in directly using the uproot library~\cite{Pivarski2019}.
It does \emph{not} require ROOT to be installed to be usable.

For long term data storage, it is also planned to implement an \enquote{archival} file format
that stores all information as text files.
Text files have the highest chance of remaining readable in the future,
as they are generally considered to be the lowest common denominator in data exchange.
Even if \ReMU (or even Python in general) should stop working at some point,
the data and response matrix could be re-used by different programs in this form relatively easily.

A publication following the response-matrix centred approach would include at least these elements:
\begin{itemize}
    \item   Response matrix binning in reco space (\enquote{reco-binning.yml})
    \item   Response matrix binning in truth space (\enquote{truth-binning.yml})
    \item   The systematically and statistically varied sparse response matrices (\enquote{response.npy})
    \item   A truth space histogram of how many events were simulated in each bin (\enquote{generator-truth.npy})
    \item   Reco histogram of data (\enquote{data.npy}) or CSV file of reco properties of all data events (\enquote{data.csv})
    \item   Optionally, truth space background templates ({background.npy})
\end{itemize}
Users of the publication could then provide their own signal predictions to calculate likelihoods.
\ReMU provides many functions to make this as easy as possible.
This includes the definition of composite hypotheses and the likelihood maximisation over their parameter spaces.

\section{Example analysis}

\newminted{python}{frame=single, gobble=4}

\subsection{Introduction}

This section is intended as a rough outline on how \ReMU and the response-matrix-centred approach could be used in practice.
Since the actual software is subject to active development,
we will concentrate on the principles rather than the actual implementation here.
The example is taken from the documentation of \ReMU,
and the reader should refer to it for the full implementation details and additional information \cite{ReMUdocs}.

The mock experiment that is handled here is quite simple.
It records events with only two properties: $x$ and~$y$.
Only~$x$ is smeared by the detector (Gaussian blur with $\sigma = 1$).
The efficiency of detecting an event depends only on~$y$ ($\epsilon = 0.9 \frac{1}{2}(1+\operatorname{erf}(y/\sqrt{2})$)).
There are no background events.
Let us assume we are interested in a measurement of the distribution of $x$.

\subsection{Preparation of the response matrix}

The response matrix must be prepared by the detector experts within an experiment's collaboration.
Only they have the necessary knowledge about the detector response and its uncertainties,
as well as access to the full detector simulation framework.
Care must be taken to ensure that the response matrix is actually as model-independent as desired.
\ReMU offers a few methods to test the matrices for that property.

Response matrix objects are created by specifying the binning in reco and truth space.
They are then filled with simulated events that were processed with the full detector simulation and analysis chain:
\begin{pythoncode}
    respA = migration.ResponseMatrix(reco_binning, truth_binning)
    respA.fill_from_csv_file("modelA_data.txt")
\end{pythoncode}

\begin{figure}
    \includegraphics[width=0.49\textwidth]{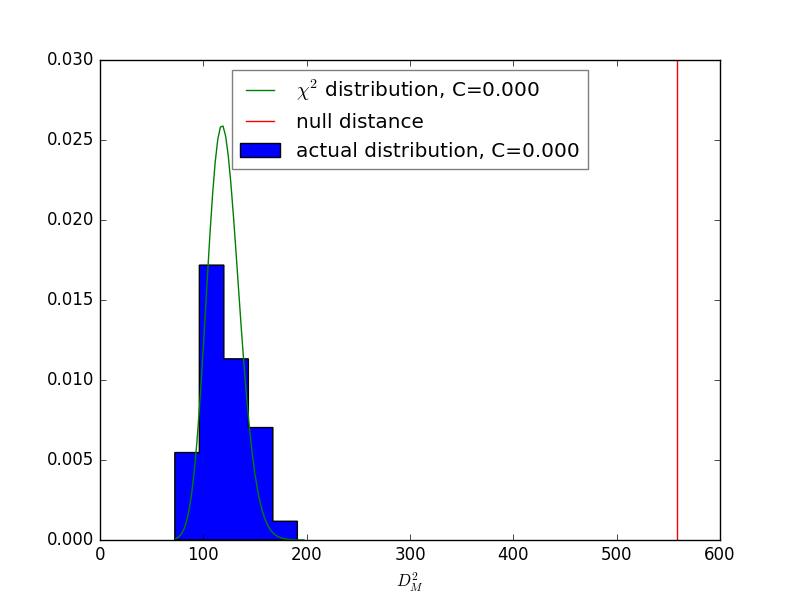}
    \includegraphics[width=0.49\textwidth]{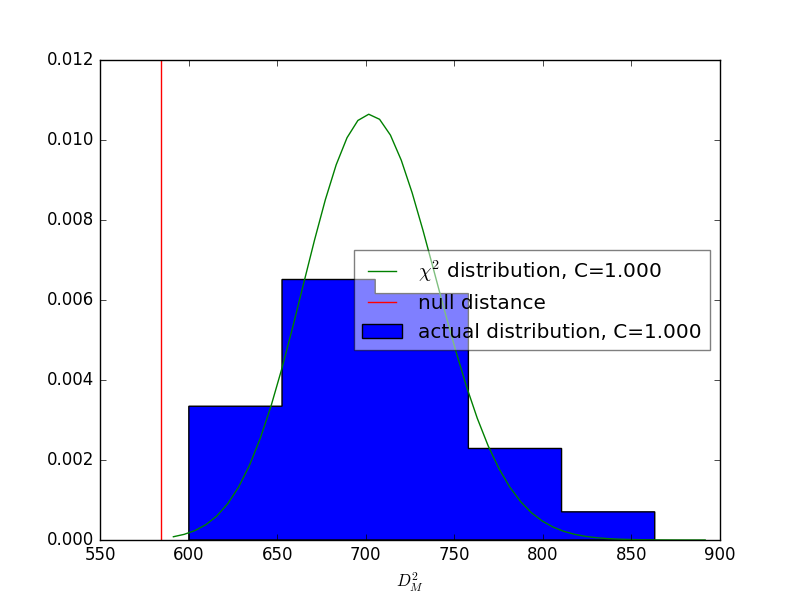}
    \caption[Matrix compatibility plots]{\label{fig:compatibility}%
        Matrix compatibility plots.
        The squared Mahalanobis distance (see \aref{sec:matrix_tests}) of two matrices populated with simulated events from different models.
        When the truth binning cannot cover the varying detector response between the models (left),
        the distance (vertical line) will be larger than expected from purely random variation (blue histogram).
        In the Gaussian limit, this variation should be chi-squared distributed (green curve).
        If the binning covers the model differences (right),
        the distance should fall within the expected distribution, or be smaller.
        The distance can be smaller than the expected distribution,
        because the parametrisation of the uncertainties of the matrix elements starts with a prior (or pseudo observations) that are common to both compared matrices (see \aref{sec:build}).
        So two matrices with no data in them will be perfectly identical,
        despite large expected statistical uncertainties.
    }
\end{figure}

A model-independent response matrix should not depend on the model that was used to populate the response matrix.
\ReMU offers a method to calculate the Mahalanobis distance between two matrices
and compare the result with the expected distribution assuming that the two matrices are random variations of the same matrix (see \aref{sec:matrix_tests}):
\begin{pythoncode}
    respB = migration.ResponseMatrix(reco_binning, truth_binning)
    respB.fill_from_csv_file(["modelB_data.txt"])
    respA.plot_compatibility("compatibility.png", respB)
\end{pythoncode}
See \aref{fig:compatibility} for how these plots might look for compatible and incompatible matrices.
Note that passing this test is a necessary condition for model-independent matrices, but not a sufficient one.
The available models for this test might be too similar to show any intrinsic model-dependencies of the response matrix.
It is up to the detector experts to make sure that they are covering the necessary response variations in the truth binning.

\begin{figure}
    \centering
    \includegraphics[width=0.8\textwidth]{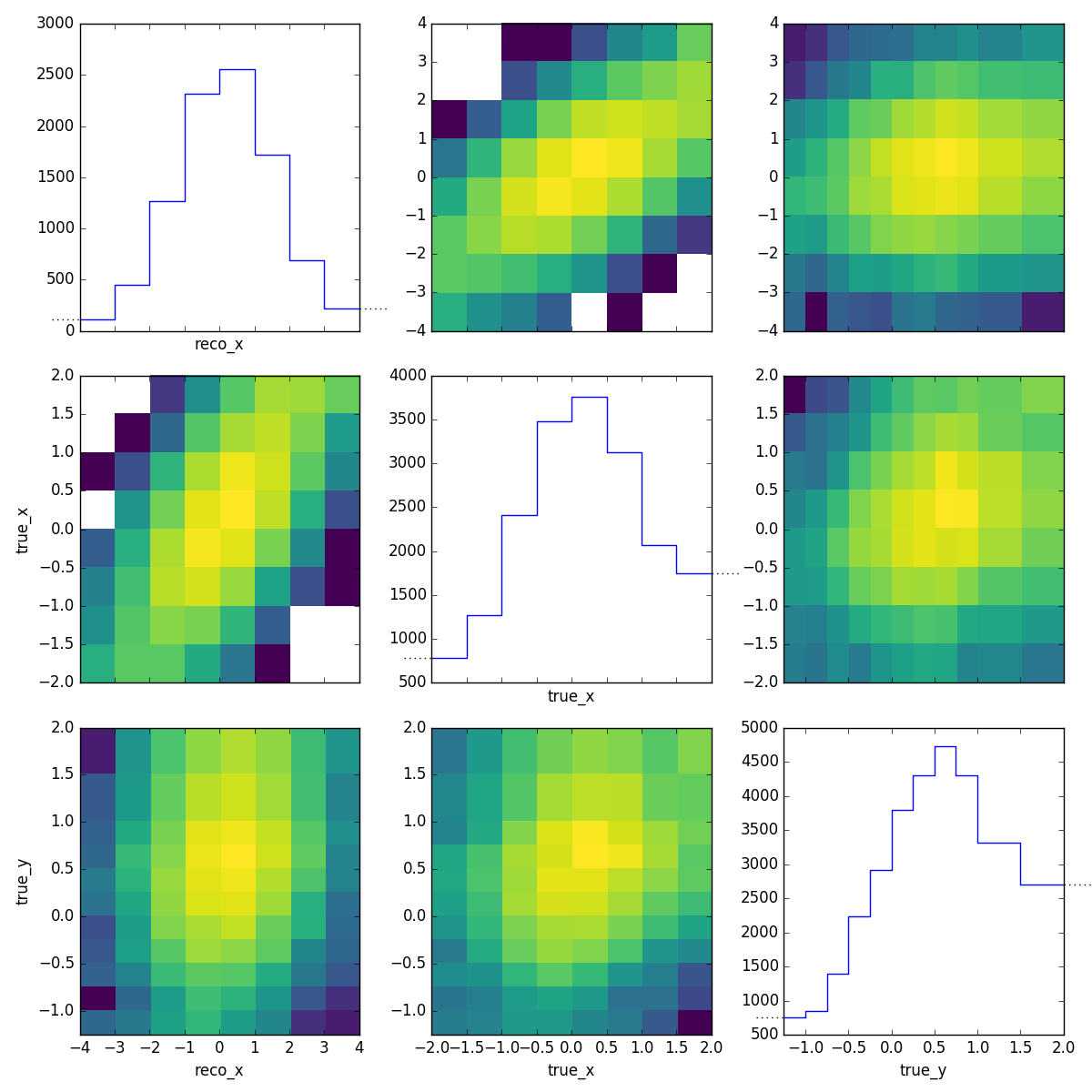}
    \caption[Response matrix projection]{\label{fig:matrix}%
        Response matrix projection.
        1D and 2D projections of the distribution of events that populate the response matrix.
        Only events that have been reconstructed are included,
        i.e. this shows the smearing/migration of events, but not the efficiency.
        Dotted lines outside the plot axes indicate that the corresponding bin is not constrained in that direction,
        i.e. it behaves like on over- or underflow bin.
    }
\end{figure}

In the case of this example, it is necessary to bin the truth both in~$x$ (because this is the variable of interest) and in~$y$ (because the detection efficiency depends on this variable).
Note that if the response matrix is model-independent, it can actually be populated by all available simulated data combined:
\begin{pythoncode}
    resp.fill_from_csv_file(["modelA_data.txt", "modelB_data.txt"])
\end{pythoncode}
\Aref{fig:matrix} shows a plot of the 2D projections of the final response matrix.
It was generated by one of the many methods in \ReMU that are intended to help gaining insights into the properties of the response matrices when building them:
\begin{pythoncode}
    resp.plot_values("optimised_response_matrix.png", variables=(None, None))
\end{pythoncode}

\subsection{Using the response matrix}

Once the response matrix (or the set of response matrices, see \aref{sec:likelihood}) is prepared and published with the data vector,
it can be used to do statistical tests.
This can be done both inside the experiment's collaboration, as well as outside of it.
The usage of the response matrix does not require expert knowledge of the detector.

Within \ReMU the data and response matrices are combined into \texttt{LikelihoodMachine} objects.
These then provide methods to do different statistical tests on model predictions.
It does not matter what exactly the data looks like or how many matrices make up the set,
the interface to the user stays the same.

The simplest kind of test that can be done is comparing the likelihoods of different models.
For this, all that is needed is a model prediction for the events in truth space:
\begin{pythoncode}
    truth_binning.fill_from_csv_file("modelA_truth.txt")
    modelA = truth_binning.get_values_as_ndarray()
\end{pythoncode}
The \texttt{LikelihoodMachine} can then calculate (log-)likelihoods of the measured data,
given this prediction and marginalising over the detector uncertainties encoded in the set of response matrices:
\begin{pythoncode}
    lm.log_likelihood(modelA)
\end{pythoncode}

In case models have free parameters, it is also possible to maximise the likelihood over the allowed parameter space.
For example, one can use the (area normalised) shape of models as templates and let the template weight (i.e. the number of true events) be fitted to the data:
\begin{pythoncode}
    modelA_shape = TemplateHypothesis([modelA / np.sum(modelA)])
    lm.max_log_likelihood(modelA_shape)
\end{pythoncode}
This will return both the maximised (log-)likelihood and the parameter values of that point.

Since likelihood values alone are hard to interpret, \ReMU also offers several methods to calculate p-values:
\begin{pythoncode}
    lm.likelihood_p_value(modelA)
    lm.max_likelihood_p_value(modelA_shape)
    lm.max_likelihood_ratio_p_value(model0, model1)
\end{pythoncode}
These respectively calculate the probability of
\begin{itemize}
    \item measuring data that yields a lower likelihood than the actual one,
        assuming the provided model is true,
    \item measuring data that yields a lower maximum likelihood than the actual one,
        assuming the best fit point of the provided model is true,
    \item measuring data that yields a lower ratio of maximised likelihoods between the two specified models than the actual one,
        assuming the best fit point of \texttt{model0} is true.
\end{itemize}
These p-values can then be used to check goodness of fit, to construct different confidence intervals, or to do frequentist hypothesis tests.

\begin{figure}
    \centering
    \includegraphics[width=0.49\textwidth]{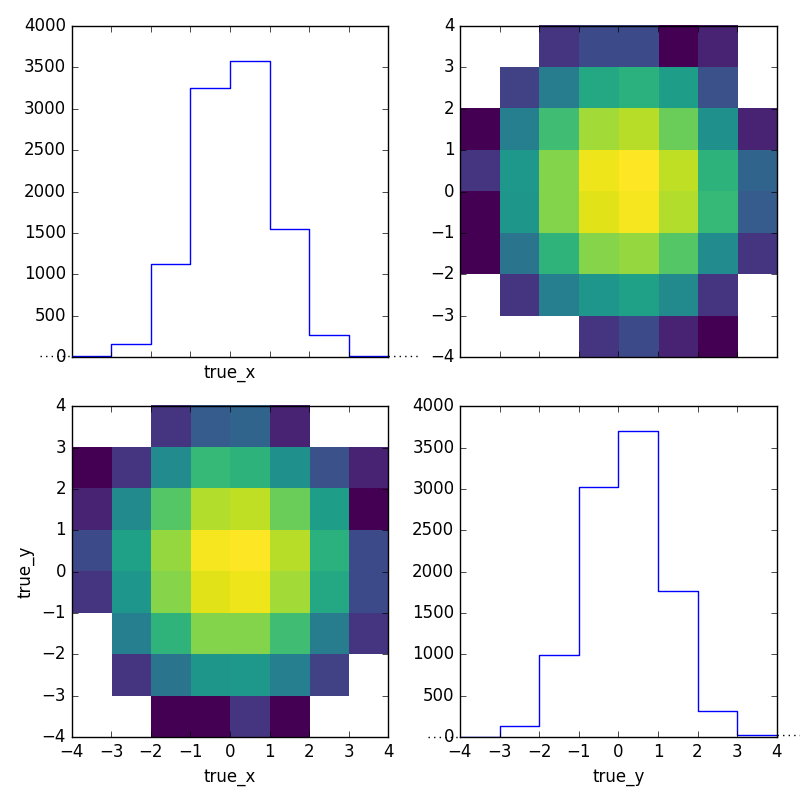}
    \includegraphics[width=0.49\textwidth]{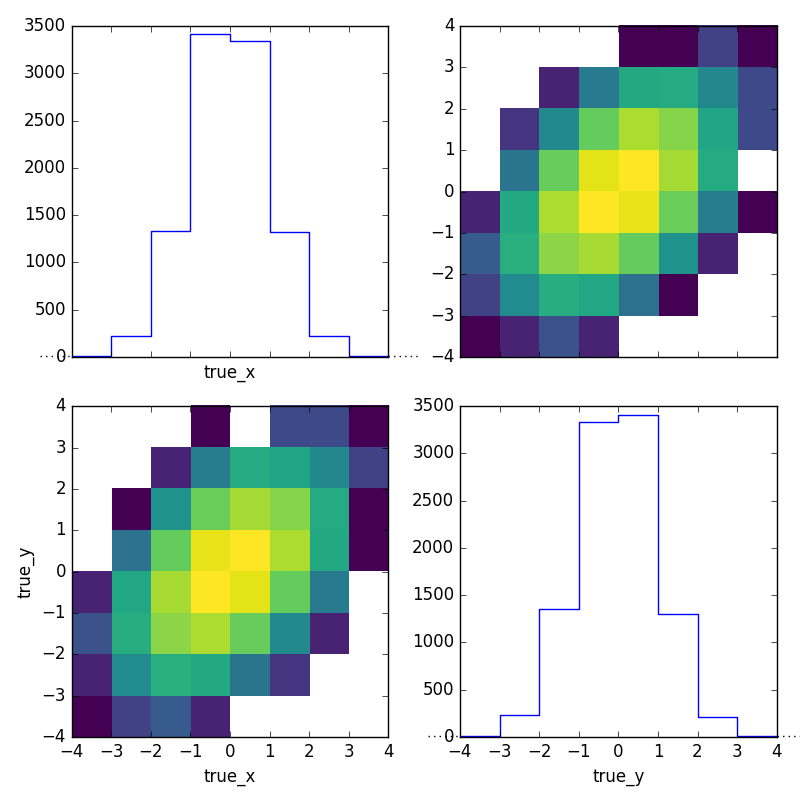}
    \caption[Example models]{\label{fig:models}%
        Example models.
        True distribution of events in model~$A$ (left) and model~$B$ (right).
        Dotted lines outside the plot axes indicate that the corresponding bin is not constrained in that direction,
        i.e. it behaves like on over- or underflow bin.
    }
\end{figure}

Let us assume we have two models we want to compare to the data.
Model~$A$ assumes that the true properties $x$ and~$y$ of events are uncorrelated, normal distributed.
Model~$B$ assumes a correlation between $x$ and~$y$ (see \aref{fig:models}).
They also feature slightly different means of the distribution.
See \aref{tab:models} for a summary of the model parameters.
Each model only predicts the shape of the event distribution, but not the total number of events.
Note that even though we are only interested in a measurement of~$x$,
the different behaviours in~$y$ lead to different average detection efficiencies between the models.

\begin{table}
    \centering
    \caption[Example models]{\label{tab:models}%
        Example models.
    }
    \begin{tabular}{lccccc}
        \tablelead{}
        & $\operatorname{E}(x)$ & $\operatorname{E}(y)$ & $\operatorname{var}(x)$ & $\operatorname{var}(y)$ & $\operatorname{cov}(x,y)$ \\
        Model $A$ & 0.1 & 0.2 & 1.0 & 1.0 & 0.0 \\
        Model $B$ & 0.0 & 0.0 & 1.0 & 1.0 & 0.5 \\
        \tabletail
    \end{tabular}
\end{table}

\begin{figure}
    \centering
    \includegraphics[width=0.5\textwidth]{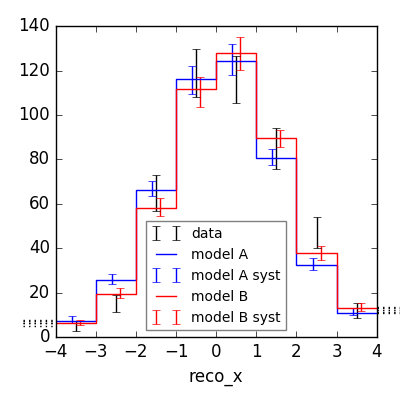}
    \caption[Reco-space model comaprison]{\label{fig:comparison}%
        Reco-space model comparison.
        The model predictions are shown with mean and standard deviation due to detector systematics.
        The data is shown with $\sqrt{N}$ \enquote{error bars} for visualisation purposes only.
    }
\end{figure}

Maximising the likelihood over the free normalisation parameter of the models yields two maximum likelihood solutions
that both fit the data reasonably well (see \aref{fig:comparison}).
A look at their respective \texttt{max\_likelihood\_p\_value}s tells us that model~$A$ is slightly disfavoured (p-value $\sim 0.1$).

\begin{figure}
    \centering
    \includegraphics[width=0.8\textwidth]{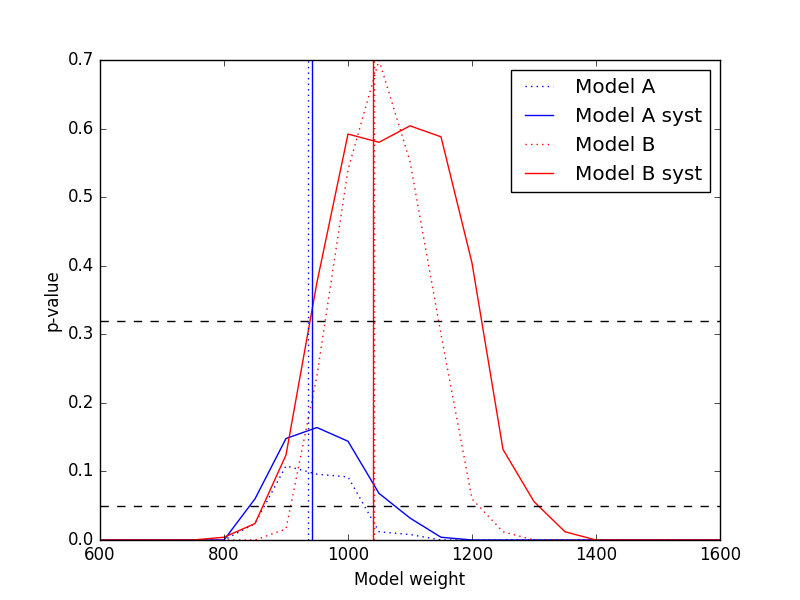}
    \caption[Local p-values]{\label{fig:local-p-values}%
        Local p-values, as a function of the template weight (number of true events).
        Vertical lines show the maximum likelihood solutions.
        The dotted lines show the results when not applying any detector systematics,
        i.e. using only a single (nominal) response matrix.
    }
\end{figure}

\begin{figure}
    \centering
    \includegraphics[width=0.8\textwidth]{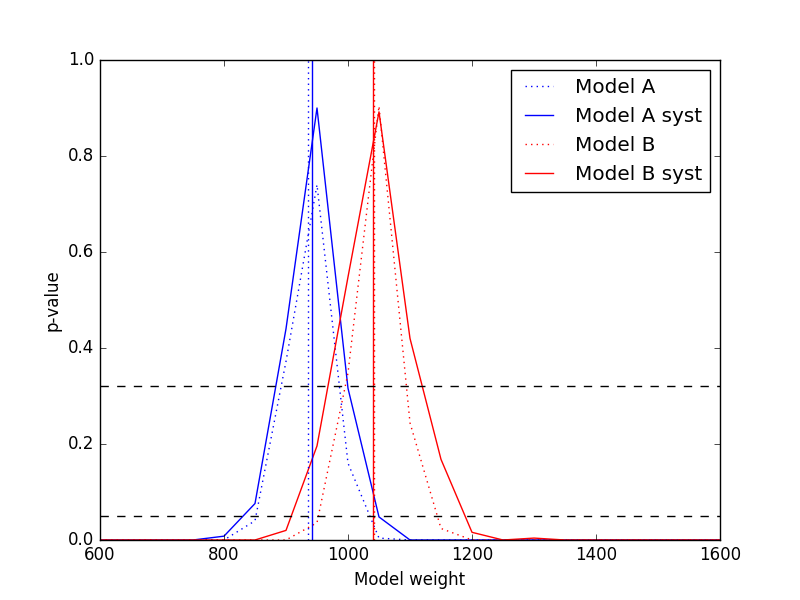}
    \caption[Likelihood ratio p-values]{\label{fig:ratio-p-values}%
        Likelihood ratio p-values, as a function of the template weight (number of true events).
        Vertical lines show the maximum likelihood solutions.
        The dotted lines show the results when not applying any detector systematics,
        i.e. using only a single (nominal) response matrix.
    }
\end{figure}

Instead of globally excluding a hypothesis, it can be useful to look at the local p-values in the parameter space.
\Aref{fig:local-p-values} shows this for the two models.
Again model~$A$ is disfavoured.
This is not useful to construct confidence intervals under the assumption that each model is true, though.
This is done in \aref{fig:ratio-p-values} using the \texttt{max\_likelihood\_ratio\_p\_value} of the fixed normalisation prediction vs the floating normalisation models.
By construction, this p-value is 1 at the maximum likelihood fit point of the parameter space.
Model~$A$ and~$B$ yield different confidence intervals for the total number of events,
because their average detection efficiencies are different.

This kind of model-dependent result is a good illustration for the advantage of the response-matrix-centred forward-folding approach of sharing data.
Had the data of this mock experiment been shared as an unfolded distribution,
it would have had to include the different average efficiencies of possible models in the systematic uncertainties of the result.
This would lead to inflated errors compared to the specific model tests done here.
But even those inflated errors could not guarantee that the coverage of the true value is as expected
if the true model is not among the considered ones.
Alternatively the result could have been unfolded in both $x$ and~$y$,
but that approach is not always feasible.
Depending on the available data statistics, the number of variables that influence the detector response, and how well the detector can measure those in the first place,
the data might not be able to constrain all relevant truth bins.
This is not an issue in the forward-folding approach.

\section{Conclusions}

The response-matrix-centred approach to presenting cross-section measurements promises to be a useful addition to the set of tools available to cross-section analysts.
It combines the fine-grained model-independence of a multi-dimensionally unfolded differential cross-section measurement with the ability to work with low-statistics, coarsely binned real data.
Even for measurements with enough real data for very fine reco-binning, the comparison in reco-space can offer superior model-separation power over comparisons in (unfolded) truth-space \cite{Cousins2016}.

A method was presented how to handle systematic detector uncertainties by providing a set of possible response matrices and calculating the marginal likelihood of their reco-space predictions.
Statistical uncertainties of the matrix elements stemming from finite Monte Carlo statistics are handled in a similar way,
by quantifying the uncertainties and creating random variations of the response matrices accordingly.

Three methods to deal with backgrounds were presented, avoiding subtracting any events from the original data vector.
Irreducible background must be added to the signal truth bins by the (background) model.
\enquote{Physics-like} background can be handled just like signal events, with its own set of truth bins and corresponding response parametrisation in the response matrix.
Detector-specific background templates can be put directly into the response matrix, with a single truth bin determining their strength.

The biggest challenge for the analyst creating the matrix is to choose an appropriate truth binning.
The detector response typically depends on a multitude of variables,
but the number of bins grows exponentially with the number of binning dimensions.
This leads to a very high demand for Monte Carlo statistics to build the response matrices.

Once the matrix is available though, it is relatively easy to test various physics models against the data,
without having to re-evaluate the detector response and its uncertainties for each model.
This will be useful for the NUISANCE\cite{nuisance} and Rivet\cite{Buckley2010} frameworks, for example.

\acknowledgments{}

I want to thank my colleagues of the T2K collaboration for supporting me during the genesis of this paper,
including -- but not limited to -- Morgan Wascko, Kendall Mahn and Stephen Dolan.
Especially Stephen has been very helpful, with many fruitful discussion about the finer points of (un-)folding, and feedback on paper drafts.
I also want to thank my colleagues at the STFC Rutherford Appleton Laboratory for their feedback and for providing a \enquote{collider physics perspective}.
This work was partially supported by the Deutsche Forschungsgemeinschaft (DFG) [grant number RO3625-1].

\printbibliography

\appendix

\section{Statistical methods}
\subsection{Simple hypotheses and absolute maximum likelihood}
\label{sec:hypothesis-tests}

A \emph{simple hypothesis} is completely characterised by the vector of truth expectation values $\bm{\mu}$.
It has no free parameters.
Each expectation value must be a non-negative real number, $\mu_j \in \mathbb{R}_{\ge 0}$.
This defines the set of all conceivable hypotheses $\Omega$:
\begin{equation}
    \Omega = \mathbb{R}_{\ge 0}^d \text{,}
\end{equation}
where $d = \dim(\bm\mu)$ is the number of truth bins.
We can thus define a maximum likelihood hypothesis $\bm{\mu}_{\text{max } L}$ such that
\begin{equation}
    \lik_\text{max}(\Omega) = \lik(\bm{\mu}_{\text{max }\lik}) = \max_{\bm{\mu} \in \Omega} \, \lik(\bm{\mu})\text{.}
\end{equation}
This hypothesis and its likelihood value can then be used as a baseline to compare other hypotheses to it.

\subsection{Likelihood ratio testing}

The agreement between the data and any given hypothesis can be evaluated with the likelihood ratio $\lambda$:
\begin{equation}
    \lambda(\bm\mu) = \frac{L(\bm\mu)}{L_\text{max}(\Omega)} \text{.}
\end{equation}
By construction, this value is in the range $[0,1]$.
A high value shows good agreement, while low values indicate disagreement.

According to the \emph{Neymanâ€“Pearson lemma}~\cite{Neyman-Pearson}, a hypothesis test using $\lambda$ as a test statistic is the most powerful%
\footnote{The power of a test $1-\beta$ describes the probability of rejecting a false hypothesis.}
test possible.
So we define a critical value $\eta$ and reject hypotheses where $\lambda < \eta$.
The choice of $\eta$ depends on the desired significance%
\footnote{The significance of a test $\alpha$ describes the probability of rejecting a true hypothesis.}
of the test $\alpha$ and the expected distribution of the likelihood ratio $f(\lambda)$ given that $\bm\mu$ is true.
It must be chosen such that
\begin{equation}
    P(\lambda < \eta | \bm\mu) = \int_0^\eta f(\lambda|\bm\mu)\,\dd\lambda \overset{!}{\le} \alpha \text{.}
\end{equation}
The distribution of $\lambda$ must be evaluated for each tested hypothesis separately,
for example by doing a sufficient number of MC experiments.
The critical value is thus a function of the hypothesis $\eta(\bm\mu)$.
This complicates direct comparisons of different hypotheses.

To get a value that is comparable between hypotheses, one can use the likelihood ratio p-value as test statistic directly.
It is the probability of measuring a likelihood \enquote{worse} than the actually measured one $\lambda_0$, assuming that the tested hypothesis is true:
\begin{equation}
    p_\lambda(\bm\mu) = P(\lambda < \lambda_0 | \bm\mu) = \int_0^{\lambda_0} f(\lambda|\bm\mu)\,\dd\lambda \text{.}
\end{equation}
By construction, this value is uniformly distributed for the true hypothesis,
so the critical value is just the significance,
and hypotheses are rejected if
\begin{equation}
    p_\lambda(\bm\mu) < \alpha \text{.}
\end{equation}

\subsection{Composite hypotheses}
\label{sec:comp-hyp}

Often a tested hypothesis will have some free (nuisance) parameters.
Those are called \emph{composite hypotheses}.
They will define the truth expectation values $\bm\mu$ as a function of these free parameters $\bm\mu(\bm\theta)$,
with the number of free parameters $d' = \dim(\bm\theta) < d$.
The possible values of $\bm\theta$ define the set of \emph{simple} hypotheses $\Theta$,
which is a subset of all conceivable hypotheses:
\begin{equation}
    \bm\mu(\bm\theta) \in \Theta \subset \Omega \quad\Leftrightarrow\quad \bm\theta \in \omega \text{,}
\end{equation}
where $\omega$ is the set of allowed values of $\bm\theta$.
For example, if all parameters are unrestricted real values, we have
\begin{equation}
    \omega = \mathbb{R}^{d'} \text{.}
\end{equation}
We consider a composite hypothesis \emph{true} if it contains the true simple hypothesis $\bm\mu_\text{true}$, and \emph{false} otherwise.

Again, we would like to test hypotheses with the highest possible power at a given significance.
To reject a composite hypothesis $\Theta$, we must reject all contained simple hypotheses $\bm\mu(\bm\theta)$:
\begin{equation}
    \lambda(\bm\mu) < \eta(\bm\mu) \quad\forall\quad \bm\mu \in \Theta \text{.}
\end{equation}
As an approximation with lower than ideal power, we can consider the maximum likelihood and
minimum critical value:
\begin{equation}
    \lambda_\text{max}(\Theta) = \max_{\bm\mu \in \Theta} \lambda(\bm\mu) \overset{!}{<} \eta_\text{min}(\Theta) = \min_{\bm\mu} \eta(\bm\mu) \text{.}
\end{equation}
Depending on the variation of $\eta(\bm\mu)$ within $\Theta$,%
\footnote{In the limit of large sample sets, the distribution of $\lambda_\text{max}$ will approach a $\chi^2$-distribution~\cite{Wilks-1938} and an exact value for $\eta$ can be chosen accordingly.}
the significance of the test will be less than or equal to the nominal value $\alpha$.

To increase the power of the test, we can use the $p$-values of the likelihood ratios directly.
With this test statistic, the critical value is identical for all simple hypotheses,
and we reject a composite hypothesis if
\begin{equation}
    p_\lambda(\bm\mu) < \alpha \quad\forall\quad \bm\mu \in \Theta \text{.}
\end{equation}
That means we can exclude a composite hypothesis by checking whether
\begin{equation}
    p_\text{max}(\Theta) = \max_{\bm\mu \in \Theta} p_\lambda(\bm\mu) < \alpha
\end{equation}
with maximum power.
This is called the \enquote{supremum method}~\cite{Demortier2008} and not computationally harder than finding $\eta_\text{min}$.
In both cases $f(\lambda|\bm\mu)$ has to be calculated for each evaluation in the minimisation/maximisation process.

\subsection{Parameter estimation}
\label{subsec:confidence}

If a composite theory $\Theta$ is not rejected, one might want to quote a set of \enquote{best fit} parameters
and/or a range of allowed values, i.e. confidence intervals.
The maximum likelihood point estimator for the parameters $\bm{\hat\theta}$ is straight forward.
It is the set of parameters that produce the highest likelihood:
\begin{equation}
    \lik(\bm\mu(\bm{\hat\theta})) = \lik_\text{max}(\Theta) \text{.}
\end{equation}

Confidence intervals for the parameters can be calculated by rejecting part of the possible parameter space analogously to the general composite hypothesis test in \aref{sec:comp-hyp}.
For this, we split the parameters into interesting parameters $\bm\theta$,
where we want to quote the intervals, and nuisance parameters $\bm\phi$.
We then interpret the set of all $\bm\mu(\bm\theta,\bm\phi)$ with a fixed $\bm\theta$ as a new composite hypothesis $\Theta(\bm\theta)$:
\begin{equation}
    \bm\mu(\bm\theta,\bm\phi) \in \Theta(\bm\theta) \subset \Theta \quad\Leftrightarrow\quad \bm\phi \in \Phi(\bm\theta) \text{,}
\end{equation}
where $\Phi(\bm\theta)$ is the set of allowed values of $\bm\phi$ given a specific $\bm\theta$.
Now we can exclude values of $\bm\theta$ by checking whether
\begin{equation}
    p_\text{max}(\Theta(\bm\theta)) < \alpha \text{.}
\end{equation}
Those values of $\bm\theta$ that have not been rejected define the confidence region.

It might be useful to construct confidence intervals for parameters of composite hypotheses whether or not they have been excluded.
In these cases, one is usually only interested in the allowed parameter range within the context of the analysed hypotheses.
This can easily be achieved by replacing the absolute maximum likelihood $\lik_\text{max}(\Omega)$ with the maximum likelihood of the hypothesis $\lik_\text{max}(\Theta)$.
The likelihood ratio is then
\begin{equation}
    \lambda(\bm\mu) = \frac{L(\bm\mu)}{L_\text{max}(\Theta)} \text{,}
\end{equation}
and the construction of the confidence interval only ever compares the nested hypothesis $\Theta(\bm\theta)$ directly with the enveloping hypothesis $\Theta$.
This can reduce the number of parameters considerably, as no evaluation of the absolute maximum likelihood needs to be done.
A lower number of free parameters decreases the computational load considerably.

\subsection{Profile plug-in p-values}
\label{sec:profile-plug-in}

Even when only comparing two hypotheses with moderate number of parameters, finding $p_\text{max}(\Theta(\bm\theta))$ is a computationally intensive task.
Calculating the p-value for a single $\bm\mu$ takes the generation of $\orderof(100)$ toy data sets from the reco predictions of that hypothesis,
and then maximising the likelihoods of both compared composite hypotheses for each data set.
Maximising the p-value with a typical optimisation algorithm means that it has to be evaluated at least $\orderof(10000)$ times, depending on the difficulty of finding the global(!) likelihood maxima.
This quickly escalates into millions upon millions necessary fits and a corresponding demand of computing power.

A drastic reduction can be achieved when using the \enquote{profile plug-in} p-value instead of the maximum p-value.
Instead of maximising the p-value over all possible hypotheses $\bm\mu \in \Theta$,
one only evaluates the p-value of the most likely hypothesis $\bm{\hat\mu}$:
\begin{equation}
    p_\text{plug}(\Theta) = p_\lambda(\bm\mu(\bm{\hat\theta})) \text{,}
\end{equation}
or in the context of parameter estimation:
\begin{equation}
    p_\text{plug}(\Theta(\bm\theta)) = p_\lambda(\bm\mu(\bm\theta,\bm{\hat\phi})) \text{.}
\end{equation}
Here $\bm{\hat\theta}$ and $\bm{\hat\phi}$ are the maximum likelihood estimates of the (nuisance) parameters:
\begin{align}
    \lik(\bm\mu(\bm{\hat\theta})) &= \lik_\text{max}(\Theta) \text{,} \\
    \lik(\bm\mu(\bm\theta,\bm{\hat\phi})) &= \lik_\text{max}(\Theta(\bm\theta)) \text{.}
\end{align}
The calculation of this value requires only a single optimisation of the likelihood.
The p-value itself is then computed with toy data assuming the truth of the estimate.

This is called \enquote{profile plug-in} p-value as we plug-in the profile maximum likelihood estimate for the nuisance values as an estimate for the distribution of the likelihood ratios of the true hypothesis.
The method has certain advantages over other approximation methods~\cite{Kabaila-2000}, but it is still an approximation.
It is thus important to check the coverage properties of any analysis using this method.

\subsection{Bayesian posterior sampling}

The exact frequentist approaches described above need a prohibitive amount of computing power when the number of parameters of the tested composite hypothesis is large.
Those models are better handled by a Bayesian approach.
Using a Markov Chain Monte Carlo (MCMC) method, it is relatively easy to sample parameter sets $\bm\theta$ from the posterior probability
\begin{equation}
    P(\bm\theta | \bm{n}) \propto L(\bm\mu(\bm\theta)) \, P(\bm\theta) \text{,}
\end{equation}
with the prior probability $P(\bm\theta)$.
These sets can then be used to infer information about the parameters, e.g. point estimates or credible intervals,
and to compare different hypotheses with one another.

Hypothesis comparisons are usually done with the \emph{Bayes factor} $K$:
\begin{equation}
    B = \frac{\int \lik(\bm\mu(\bm\theta_0)) P(\bm\theta_0)  \dd \bm\theta_0}{\int \lik(\bm\mu(\bm\theta_1)) P(\bm\theta_1)  \dd \bm\theta_1} \text{,}
\end{equation}
i.e. the ratio of prior mean likelihoods.
To Bayesian posterior odds $K$ are then just the Bayes factor multiplied by the prior odds:
\begin{equation}
    K = B \frac{P(\Theta_0)}{P(\Theta_1)} \text{.}
\end{equation}
When averaging the likelihood over the posterior distributions of the parameters,
one gets the \emph{posterior Bayes factor} $B_\text{post}$ \cite{Aitkin1991}:
\begin{equation}
    B_\text{post} = \frac{\int \lik(\bm\mu(\bm\theta_0)) P(\bm\theta_0|\bm{n})  \dd \bm\theta_0}{\int \lik(\bm\mu(\bm\theta_1)) P(\bm\theta_1|\bm{n})  \dd \bm\theta_1} \text{.}
\end{equation}
Using the posterior Bayes factor reduces the influence of the priors of the parameters on the result,
but it is sometimes criticised  for \enquote{using the data twice}:
once for determining the posterior distributions of the parameters $P(\bm\theta|n)$
and once for calculating the likelihood $\lik(\bm\mu(\bm\theta))$ to be averaged over those distributions.

Another possible method is using the \emph{Posterior distribution of the Likelihood Ratio} (PLR) to infer the data preference of one model over another.
The PLR is defined as the posterior probability of the likelihood ratio of the compared hypotheses being below or equal to a certain threshold value:
\newcommand{\PLR}{\mathrm{PLR}}
\begin{equation}
    \PLR_{\Theta_0, \Theta_1}(\bm{n}, \zeta) = P\left(\frac{\lik(\bm\mu(\bm\theta_0))}{\lik(\bm\mu(\bm\theta_1))} \le \zeta \,\middle|\, {\bm\theta_0 \sim P(\bm\theta_0 | \bm{n}), \bm\theta_1 \sim P(\bm\theta_1 | \bm{n})} \right) \text{,}
\end{equation}
with $\bm\mu(\bm\theta_0) \in \Theta_0$ and $\bm\mu(\bm\theta_1) \in \Theta_1$ the (completely independent) parametrisations of the tested hypotheses.
If the threshold value $\zeta$ is set to 1, the PLR is equivalent to a frequentist p-value under certain circumstances~\cite{Ferrari-2014}.
But even when this is not the case, the interpretation is straight forward:
$\PLR_{\Theta_0, \Theta_1}(\bm{n}, \zeta=1)$ is the posterior probability of the data being more likely under $\Theta_1$ than under $\Theta_0$.

For all Bayesian analyses, it is important to choose suitable priors $P(\bm\theta)$.
There is no single \enquote{correct} way to do this,
but one useful \enquote{non informative} prior is the \emph{Jeffreys prior}~\cite{Jeffreys-1946}.
Its main advantage is that its probability density -- and especially the posterior probability density resulting from using this prior --
is \emph{invariant} under variable transformations.
This means that the results of the analysis do \emph{not} depend on the particular parametrisation of $\bm\mu(\bm\theta)$.
A drawback of Jeffreys priors is that they are not necessarily proper, i.e. they cannot always be normalised.
This is not necessarily a problem here though, as long as the posterior is well defined.

The Bayesian approach treats all unknown parameters equal.
It is therefore natural to also include the detector uncertainties in the MCMC sampling.
We simply treat the detector toy index as additional (nuisance) parameter of the model.
The posterior probability thus also includes information about how likely or unlikely the different toy detectors are:
\begin{align}
    P(\bm\theta, t | \bm{n}) &\propto L^t(\bm\mu(\bm\theta)) \, P(\bm\theta, t) \nonumber \\
        &= P(\bm{n} | \bm\mu(\bm\theta), R^t) \, P(\bm\theta, t) \text{.}
\end{align}

\end{document}